\documentclass[aps,reprint,superscriptaddress]{revtex4-2} 

\usepackage[utf8]{inputenc}
\usepackage{natbib}
\usepackage{amsmath, amssymb}

\usepackage{graphicx}
\graphicspath{{./figs/}}
\usepackage[textwidth=7.24in,top=1in,bottom=1in]{geometry}

\usepackage[unicode]{hyperref}
\hypersetup{
	colorlinks=true,
	citecolor=blue,
	linkcolor=black,
	urlcolor=blue,
}

\newcommand{\reffig}[2][]{Fig.~\ref{#2}\MakeUppercase{#1}}
\newcommand{\refeq}[1]{Eq.~(\ref{#1})}

\newcommand{\reffigS}[2][]{Fig.~S#2\MakeUppercase{#1}}
\newcommand{\reftabS}[1]{Table~S#1\MakeUppercase}
\newcommand{\refeqS}[1]{Eq.~(S#1)}

\newcommand{\kex}{\kappa_\mathrm{ex}}
\newcommand{\ki}{\kappa_\mathrm{i}}

\newcommand{\Ej}{E_\mathrm{J}}
\newcommand{\Ec}{E_\mathrm{C}}
\newcommand{\Eg}{E_\mathrm{g}}

\newcommand{\SM}{Supplementary Materials}

\newcommand{\adag}[1]{\hat{a}_{#1\vphantom{l}}^\dag}
\newcommand{\anni}[1]{\hat{a}_{#1\vphantom{l}}^{\vphantom{\dag}}}
\newcommand{\fnoise}[1]{\hat{f}_{#1\vphantom{l}}^{\vphantom{\dag}}}
\newcommand{\omeg}[2][]{\omega_{#2\vphantom{l}}^{#1\vphantom{\dag}}}
\newcommand{\fany}[2][]{f_{#2\vphantom{klmnro}}^{#1\vphantom{\prime}}}
\newcommand{\fro}{\fany{\mathrm{ro}}}

\newcommand{\chii}[2][]{\chi_{#2\vphantom{j}}^{#1\vphantom{+}}}

\begin{document}
\title{Nonequilibrium plasmon liquid in a Josephson junction chain}
\author{Anton V. Bubis}
\affiliation{Institute of Science and Technology Austria, Am Campus 1, Klosterneuburg, 3400, Austria}

\author{Lucia Vigliotti}
\affiliation{Institute of Science and Technology Austria, Am Campus 1, Klosterneuburg, 3400, Austria}

\author{Maksym Serbyn}
\affiliation{Institute of Science and Technology Austria, Am Campus 1, Klosterneuburg, 3400, Austria}

\author{Andrew P. Higginbotham}
\email{ahigginbotham@uchicago.edu}
\affiliation{James Franck Institute and Department of Physics, University of Chicago, 929 E 57th St, Chicago, Illinois 60637, USA}
\affiliation{Institute of Science and Technology Austria, Am Campus 1, Klosterneuburg, 3400, Austria}

\begin{abstract}
Equilibrium quantum systems are often described by a gas of weakly-interacting normal modes.
Bringing such systems far from equilibrium, however, can drastically enhance mode-to-mode interactions.
Understanding the resulting liquid is a fundamental question for quantum statistical mechanics, and a practical question for engineering driven quantum devices.
To tackle this question, we probe the nonequilibrium kinetics of one-dimensional plasmons in a long chain of Josephson junctions.
We introduce multimode spectroscopy to controllably study the departure from equilibrium, witnessing the evolution from pairwise coupling between plasma modes at weak driving to dramatic, high-order, cascaded couplings at strong driving. 
Scaling to many-mode drives, we stimulate interactions between hundreds of modes, resulting in near-continuum internal dynamics.
Imaging the resulting nonequilibrium plasmon populations, we then resolve the non-local redistribution of energy in the response to a weak perturbation -- an explicit verification of the emergence of a strongly interacting, non-equilibrium liquid of plasmons.
\end{abstract}

\maketitle

\section*{Introduction}
In wave systems, nonlinearities cause the transfer of energy and momentum between resonant modes.
Long of interest in studies of solids \cite{peierls_annalen_1929}, classical thermalization \cite{fermi_alamos_1955}, and wave turbulence \cite{nazarenko_book_2011}, coupled multimode dynamics has more recently gained attention in engineered quantum systems due to the development of high-quality multimode resonators in several experimental platforms.
Indeed, the applications of multimode resonators in quantum science are diverse, including non-Hermitian mechanics \cite{xu_nature_2016,patil_nature_2022}, signal processing \cite{ruesink_ncomms_2018,sivak_prapp_2020}, memory \cite{naik_ncomms_2017,hann_prl_2019,matanin_prapp_2023}, quantum thermalization \cite{pekola_prr_2024}, and many-body extensions of cavity quantum electrodynamics \cite{sundaresan_prx_2015,moores_prl_2018,guo_prl_2019}.

Long chains of Josephson junctions are a particularly pristine realization of a one-dimensional multimode quantum system.
The Josephson inductance results in a low speed of light, allowing the realization of chip-scale microwave multimode resonators.
Josephson chains have long been studied as realizations of quantum $XY$ models~\cite{haviland_1996_prb,chow_prl_1998,pop_natphys_2010,ergul_njp_2013}, undergoing a superconductor-insulator quantum phase transition due to a competition of Josephson and charging energies \cite{bradley_prb_1984,choi_prb_1998,fazio_physrep_2001}.
In the microwave domain, Josephson chains are useful for quantum-limited amplification \cite{yurke_apl_1996,castellanos-beltran_natphys_2008,bergeal_phase-preserving_2010}, and super-inductance \cite{manucharyan_science_2009,masluk_prl_2012}.

More recently, microwave spectroscopy of long chains has been used as a means to understand both superconductor-insulator \cite{kuzmin_natphys_2019,mukhopadhyay_natphys_2023}, and impurity problems~\cite{leger_natcommun_2019,mehta_nature_2023,crescini_natphys_2023}.
Despite this progress, the frontier of high energy-density, where nonlinearities result in strong mode-to-mode interactions and the emergence of a plasmon liquid, remains open.
Here and throughout, we use the term plasmon liquid in the context of nonlinear optics and wave physics, to denote strongly interacting, collective dynamics of bosonic excitations rather than the presence of a quadratic (massive) dispersion. 
In our system, the plasmons retain a gapless dispersion without a mass, similar to the hydrodynamic regimes of light and surface plasmons described in \cite{lin_prl_2013}, rather than the massive polariton fluids found in confined cavities \cite{carusotto_quantum_2013}.
What are the key internal dynamical processes of this plasmon liquid, and how can they be cleanly observed and characterized? 

Here, we address these questions by implementing multimode microwave spectroscopy to investigate the internal dynamics of Josephson junction chains.
The discrete set of plasma mode excitations in the Josephson junction chains are described by an effective bosonic Hubbard model,
\begin{equation} \label{eq:hamiltonian-full}
	H = H_0+H_\mathrm{int} = \sum_{k>0} \hbar \omega_k \adag{k} \anni{k} +  \sum_{klmn} \hbar K_{klmn} \adag{k} \adag{l} \anni{m} \anni{n},
\end{equation}
where $\adag{k}$ and $\anni{k}$ are creation and annihilation operators of given plasma mode $k$.
The first term $H_0$ captures energies of individual modes, see~\reffig{fig:1}, corresponding to excitation of individual standing waves that are only weakly interacting in equilibrium.
The second interaction term, $H_\mathrm{int}$ correspond to momentum and energy-conserving multi-mode scattering processes described by matrix element $K_{klmn}$.

The Hamiltonian~(\ref{eq:hamiltonian-full}) provides a generic description of nonlinear multimode cavity, and is relevant to a broad range of physical systems.
Previous work in Josephson systems has focused on parametrically pumping isolated terms in $H_\mathrm{int}$, including two-mode squeezing \cite{yurke_apl_1996,castellanos-beltran_natphys_2008,bergeal_phase-preserving_2010,eichler_quantum-limited_2014,sivak_prapp_2020,jolin_multipartite_2023} and beam-splitter interactions \cite{zakka-bajjani_quantum_2011}.
In contrast, this work seeks to understand the driven dynamics in the many-body setting where a large number of terms in $H_\mathrm{int}$ are resonant and compete with wave dispersion. 
Another interesting point of comparison is one-dimensional electronic systems \cite{haldane_luttinger_1981,imambekov_rmp_2012}, where nonlinear Luttinger plasmons described by \refeq{eq:hamiltonian-full} been observed in several systems \cite{desphande_electron_2010,barak_interacting_2010,wang_nonlinear_2020}.
From the Luttinger liquid perspective, our work opens new avenues by directly measuring the predicted quasiparticle lifetime of plasmons \cite{lin_prl_2013} in an engineered superconducting circuit.
From the superconducting circuits perspective, the quasiparticle lifetime of plasmons is a many-body extension of the known nonlinear damping phenomena in a nonlinear Kerr oscillator \cite{yurke_apl_1996,gely_apparent_2023}.
We emphasize that the effects mentioned above, including the internal plasmon lifetime that we observe, can be largely understood from the perspective of classical nonlinear physics.

The first question we address is the origin of the nonlinearity in \refeq{eq:hamiltonian-full}, as it is \textit{a-priori} unclear if it will be dominated by the typical gradient expansion of the Josephson potential, or rather by quantum phase slips -- events where the superconducting phase tunnels by $2 \pi$ \cite{bard_prb_2018}.
By weakly driving individual nonlinear terms, we determine that the matrix elements in \refeq{eq:hamiltonian-full} increase with wavenumber, signaling that quantum phase slips are negligible in the weak-driving regime.
Further increasing the drive strength, we observe key features of energy and momentum redistribution in nonlinear, one-dimensional quantum systems: cascaded couplings between modes compete with wave dispersion, resulting in strongly hybridized clusters of plasma modes.
High mode occupations created by driving result in a continuous broadening of the plasma modes due to momentum-conserving scattering, which we quantify by extending the kinetic framework for thermal transport in one-dimensional quantum liquids~\cite{imambekov_rmp_2012,lin_prl_2013,bard_prb_2018,pekola_prr_2024}.
For the strongest drives, we observe unexpectedly large decay rates of low wave-number modes, hinting at the emergence of phase slips in a strongly non-equilibrium regime.
Our findings establish Josephson junction chains as a versatile platform for investigating multimode quantum dynamics, with broader implications for one-dimensional quantum systems and circuit quantum electrodynamics.

\section*{Results}
\subsection*{Model system and multimode spectroscopy}

\begin{figure*}
	\centering
	\includegraphics[width=4.76in]{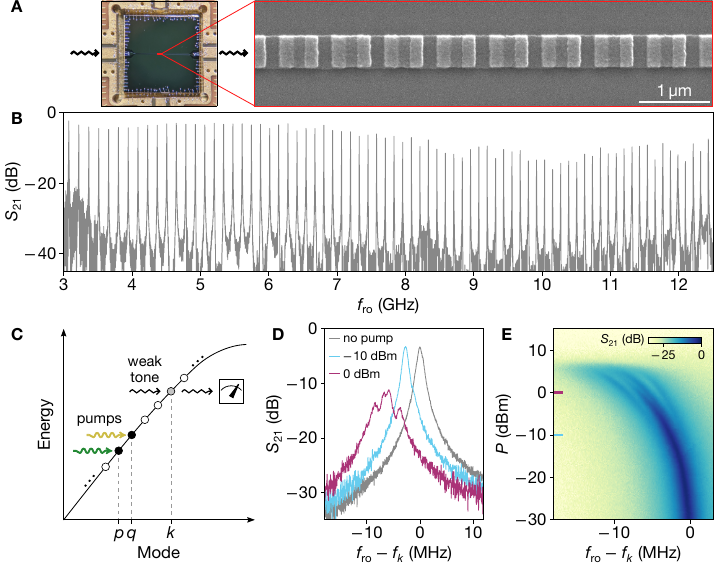}
	\caption{%
	\textbf{Microwave spectroscopy of the JJ-chain.}
	\textbf{(A)}~An optical micrograph of the device mounted on a sample holder, with a silicon chip measuring $7\!\times\!7$~mm.
	To the right, a scanning electron microscopy (SEM) image of the JJ-chain is shown.
	\textbf{(B)}~Linear response transmission magnitude ($S_{21}$) of the device with background cross-talk subtracted (see \reffigS{4}).
	\textbf{(C)}~Schematic of the three-tone measurement: two coherent tones pump modes $p$ and $q$, while a weak read-out tone near mode $k$ is used to measure $S_{21}$.
	\textbf{(D-E)}~$S_{21}$ data measured around mode $k=29$ with modes $p=25$ and $q=26$ pumped at equal power $P$ at room temperature.
	In~\textbf{(D)}, $S_{21}$ without pumps is plotted in gray for comparison.%
	}
	\label{fig:1}
\end{figure*}

We study a 5 mm-long chain of Josephson junctions as a model for a nonequilibrium quantum many-body system.
The sample is fabricated using standard electron-beam lithography and double-angle shadow evaporation of Al/AlO$_\mathrm{x}$/Al junctions (\reffig[a]{fig:1}).
Both ends of the chain are connected to microwave lines within a dilution refrigerator, enabling measurements of transmission and reflection (see setup details in \reffigS{1}).
All measurements are performed at a base temperature of 12~mK.
The linear response transmission, $S_{21}$, as a function of the read-out frequency $\fro$, reveals a dense series of transmission peaks at frequencies $\fany{k}$ (\reffig[b]{fig:1}), corresponding to the plasma modes of the chain in \refeq{eq:hamiltonian-full}~\cite{masluk_prl_2012}, $\fany{k}\approx\omega_k/(2\pi)$, where the deviation stems from the self-energy correction due to Kerr effect~\cite{krupko_prb_2018,krupko_erratum_2023}.
Although the chain is galvanically coupled to the microwave launchers, boundary conditions defining discrete set of modes arise naturally due to the impedance mismatch between the 50~$\Omega$ network and the chain, which has a high characteristic impedance of $Z > 10$~k$\Omega$ (see \reftabS{1}), typical for superinductors~\cite{kuzmin_natphys_2019}.
Plasma modes are labeled by an integer number $k=1,\ldots$ and can be though as counter-propagating standing waves with quasimomenta $\propto \pm k$.
Consequently, two mode scattering processes $m, n\to k,l$ in \refeq{eq:hamiltonian-full} conserve momenta modulo sign, $|m|+|n|= |k|+|l|$.

A first glimpse of the response to multimode drives is given by exciting two modes, $p$ and $q$ (here and below $p<q$) with pump tones, while measuring a third mode, $k$, with a weak, linear response read-out tone (\reffig[c]{fig:1}).
For low pump-tone power, the read-out mode experiences a Kerr-like frequency shift arising from the self-energy correction terms in $H_\mathrm{int}$, see \refeq{eq:hamiltonian-full}, while the transmission peak retains its Lorentzian shape~\cite{masluk_prl_2012,weissl_prb_2015,krupko_prb_2018}.
At elevated pump powers the transmission peak splits, with up to four distinct peaks resolved (\reffig[d]{fig:1}).
This splitting smoothly emerges with increasing pump power $P$, see \reffig[e]{fig:1}.
Below, we show that the observed non-Lorentzian peak shape originates from interaction term in \refeq{eq:hamiltonian-full}, highlighting fragility of the individual modes with respect to intermode interactions.
By tuning the driving, we explore the effect of interactions and the fate of modes under weak, strong, and many-mode driving conditions.

\begin{figure*}
	\centering
	\includegraphics[width=7.24in]{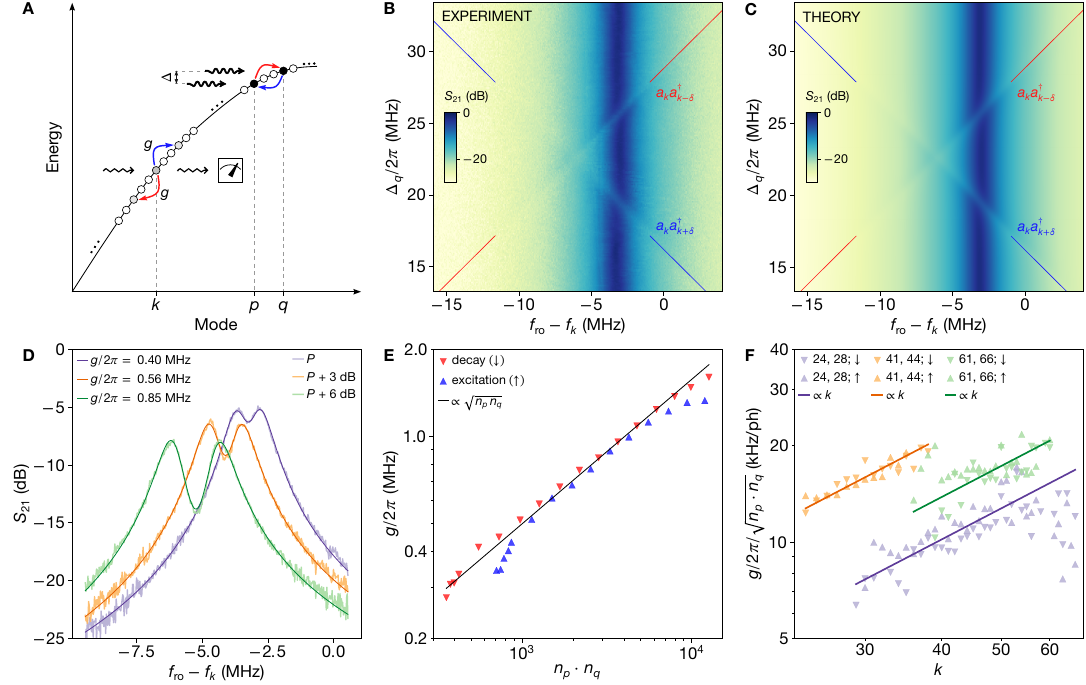}
	\caption{%
	\textbf{Matrix element scaling.}  
	\textbf{(A)}~The mode of interest $k$ couples to modes $k\pm\delta$ with the coupling strength $g$.
	By sweeping the detuning between two pumps, $\Delta/2\pi$, resonant coupling can be achieved.
	\textbf{(B, C)}~Experimental and theoretical transmission around mode $k=29$, when modes $p=41$ and $q=44$ are pumped.
	The colored guides indicate energy-conserving couplings between modes $k$ with $k+\delta$ (blue) and $k$ with $k-\delta$ (red).
	\textbf{(D)}~Transmission for resonant coupling between $k$ with $k-\delta$ for different pump powers.
	Darker lines are fits of the experimental data to the equation \refeqS{25}.
	Pump detuning dependence for these pump powers is shown in \reffigS{7}.
	\textbf{(E)}~Coupling strength $g$ vs product of occupations of pumped modes, $n_p n_q$, closely follows a square root law.
	Different markers represent decay-like coupling (modes $k$ and $k-\delta$) and excitation-like coupling (modes $k$ and $k+\delta$).
	\textbf{(F)}~Coupling strength $g$, normalized by the occupations of the pumped modes, increases with the read-out mode $k$.
	In each dataset pumped modes are fixed and both decay-like and excitation-like couplings are measured (see legend).%
	}
	\label{fig:2}
\end{figure*}

In a weak-drive regime the read-out mode is resonantly coupled only to its two energy and momentum-conserving ``nearest-neighbor'' modes (\reffig[a]{fig:2}).  
The notion of nearest-neighbor modes emerges from the fact that a given mode $k$ can now borrow or give away momentum difference of  $\delta = q-p$, where $p$ and $q$ are numbers of the pumped modes.
Thus, mode $k$ can couple to modes $k \pm \delta$.
Energy conservation imposes the additional constraint of the frequency difference between pumps $\Delta/2\pi$ matching the energy difference between modes $k$ and $k \pm \delta$.
Here the curvature of the dispersion relation $\omega_k$ in \refeq{eq:hamiltonian-full} is crucial: $\fany{k+\delta} - \fany{k} \ne \fany{k} - \fany{k-\delta}$, see~\reffig[c]{fig:1}, thus in general modes $k$ with $k-\delta$ and $k$ with $k+\delta$ are coupled resonantly for different $\Delta$.  
Experimentally, only one pump is detuned from the bare (undriven) mode frequency with the detuning $\Delta_{q} / 2\pi>0$ and $\Delta_{p} =0$  or vice versa to prevent bifurcation~\cite{eichler_epj-qt_2014,muppalla_prb_2018,andersen_pra_2020}.

In \reffig[b]{fig:2}, transmission is measured around mode $k$ while sweeping pump detuning of the mode $q$.  
The measurement reveals additional features in the lineshape of mode $k$, manifested as two straight lines in \reffig[b]{fig:2}.
The occurrence of additional features in the lineshape of mode $k$ is consistent with effect of the interaction terms from \refeq{eq:hamiltonian-full} restricted to modes $p$, $q$, $k$ and $k\pm \delta$ due to momentum and energy conservation:
\begin{equation}
	H'_\mathrm{int} = \hbar K \left[\adag{p}\anni{q}\adag{k+\delta}\anni{k} + \adag{q}\anni{p}\adag{k-\delta}\anni{k} + \mathrm{h.c.}\right].
	\label{eq:hamiltonian}  
\end{equation}  
In the presence of strong drives, standard input-output theory and linearization techniques allow replacing operators of driven modes $q$ and $p$ by their expectation values, leading to effective beam-splitter-like interaction $g\adag{k\pm\delta}\anni{k}$ (see \SM). 

At a qualitative level, the beam-splitter interaction enables hybridization of modes $k$ and $k\pm \delta$ that modifies the lineshape.
The conservation of energy in the process of scattering of photon from mode $k+\delta$ to the vicinity of mode $k$ mediated via pumped modes gives the condition $\fany[\prime]{k+\delta}-\fro = \Delta/2\pi$, where $\fany[\prime]{k+\delta}$ is the frequency of mode $k+\delta$ for particular drive (here and below with prime we emphasize the Kerr shift of mode frequency due to pumps).
For range of detunings of pump $q$ where $\fany[\prime]{k+\delta}$ remains nearly constant, this condition parametrizes a line with slope $-1$, along which the feature is located.
Analogously, scattering from mode $k-\delta$ gives a feature located along the line with slope $+1$, see~\reffig[b]{fig:2}.
At the intersections of these lines with the resonance of mode $k$, two avoided crossings occur.
At the quantitative level, an analytical prediction for transmission $S_{21}$ based on $H'_\mathrm{int}$ is shown in \reffig[c]{fig:2}.
The theoretical description uses just two cuts at fixed value of $\Delta_q$ from the experimental data to fit matrix elements, $f'_k$, and linewidths, and shows remarkable agreement with experiment (for details see \SM).

We now turn to the examination of the experimentally extracted mode coupling  strength $g$ of the effective interaction $g\adag{k\pm\delta}\anni{k}$.  
For the interaction \refeq{eq:hamiltonian}, in the limit of classical pumps of modes $p$ and $q$ with occupations $n_p$ and $n_q$, the coupling rate is proportional to the matrix element $K$ and the geometric mean of occupations: $g = |K| \sqrt{n_p n_q}$ (see \SM).  
In \reffig[d]{fig:2}, a few measured transmission spectra of the avoided crossing are plotted together with their fits for different pump powers.  
The coupling $g$ increases as a function of $P$, but slower than $g \propto P$ because, as pump power increases, modes $p$ and $q$ experience Kerr shifts, thus receiving fewer photons.  
To account for this effect, for each pump configuration, modes $p$ and $q$ were measured in reflection to extract the mode frequencies, as well as the internal, $\ki$, and external, $\kex$, couplings.  
This allows the calculation of occupations $n_p$ and $n_q$~\cite{aspelmeyer_rmp_2014}, and the results of this analysis are plotted in \reffig[e]{fig:2} for both avoided crossings, where $g$ was extracted by fitting $S_{21}$ as in \reffig[d]{fig:2}.
The data follow the $g \propto \sqrt{n_p n_q}$ trend, confirming that \refeq{eq:hamiltonian-full} is the correct starting point, and allowing the matrix element $K/2\pi = 16$~kHz/ph to be extracted, which is in reasonable agreement with the expected value of $5~\mathrm{kHz/ph}$ based on our independently determined device parameters and cryogenic insertion loss.

Next, the dependence of $g$ on the read-out mode $k$ was investigated (\reffig[f]{fig:2}).
The experimental $S_{21}$ data for the reported combinations of $p,q,k$ were fit using the same procedure as in \reffig[d]{fig:2}.
For each configuration, the pump power was chosen to remain in the regime $g \lesssim\kappa$, where $\kappa$ is the total linewidth, which offers a good trade-off between fit quality and the increasing deviation from the observed scaling $g \propto \sqrt{n_pn_q}$ at higher powers.  
For fixed modes being pumped, $g$ increases as the read-out mode number $k$ is increased, and corresponding power law is close to $g \propto k$, giving insights into the microscopic mechanism responsible for the interaction term in \refeq{eq:hamiltonian-full} as follows.
Josephson potential gives nonlinearity via higher order (gradient) expansion terms, with the matrix element that increases as $g \propto \sqrt{k \cdot (k\pm\delta)}\propto k$~(see \SM).
The competing contribution to scattering from quantum phase slips -- excitations that correspond to sudden winding of the phase by $2\pi$ -- is expected to decrease with mode frequency and $k$~\cite{bard_prb_2018,houzet_prl_2019,burshstein_prxquatum_2024}.
The experimentally observed dependence indicates that the gradient nonlinearity, as opposed to phase slips, is dominant at weak drives.

\subsection*{Nonequilibrium photon cascades}

\begin{figure*}
	\centering
	\includegraphics[width=4.76in]{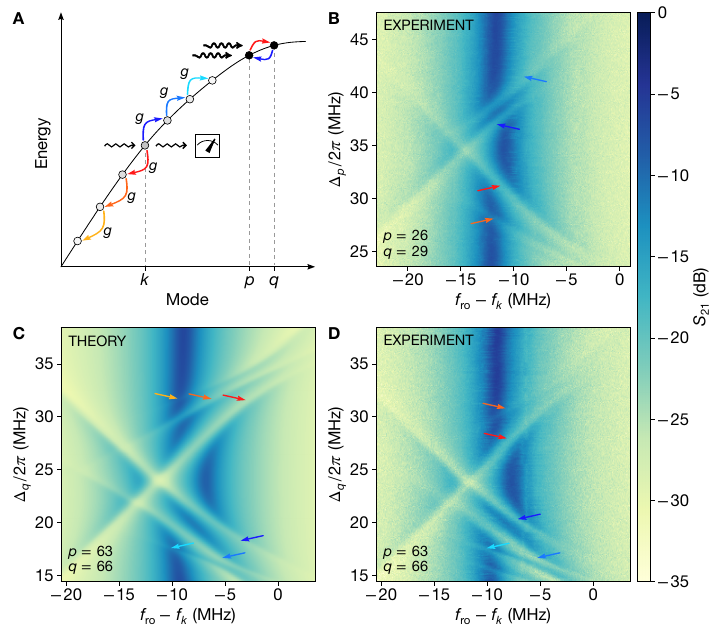}
	\caption{%
	\textbf{Cascades.}
	\textbf{(A)}~For sufficiently strong drives, mode $k$ couples to modes separated by integer multiples of $\delta$, $i\cdot\delta$, where $i$ takes both positive and negative integer values.
	\textbf{(B, D)}~Transmission measured around mode $k=46$ with two strong drives applied.
	Additional features appear for both pumping configurations: $p,q < k$ \textbf{(B)} and $p,q > k$ \textbf{(D)}.
	\textbf{(C)}~Numerically calculated transmission for the same configuration as in~\textbf{(D)}.
	In~\textbf{(B-D)}, colored arrows indicate processes schematically depicted in~\textbf{(A)}, where blue arrows represent excitation-like processes and red arrows represent decay-like processes.%
	}
	\label{fig:3}
\end{figure*}

After establishing multimode spectroscopy in the weak-driving regime and revealing the microscopic mechanism of nonlinearity in the effective Hamiltonian~(\ref{eq:hamiltonian-full}), we consider the more complex case of stronger drives illustrated in \reffig[a]{fig:3}.
Stronger pumping of two modes $p$ and $q=p+\delta$ enables coupling of an observed mode $k$ to multiple further ``neighbors'' in a cascaded fashion: mode $k$ couples to $k \pm \delta$, which in turn couples to $k \pm 2\delta$, and so on.
To describe this cascaded coupling, we use the following notation: the coupling of mode $k$ with mode $k + i \delta$ is referred to as the $i$-th order process, where $i$ can be both positive and negative.
Using similar energy conservation argument as above, we deduce that $i$-th order processes require contribution from $i$ photon scattering between pumped modes, resulting in a condition $\fany[\prime]{k+i\delta}-\fro = i \cdot \Delta/2\pi$.
For our pumping configuration where $\fany[\prime]{k+i\delta}$ stays nearly constant this condition parametrizes the line with slope of $-1/i$ for $p,q > k$ in ($\fro-\fany{k}$, $\Delta_q/2\pi$) coordinates or $1/i$ for $p,q < k$ in ($\fro-\fany{k}$, $\Delta_p/2\pi$) coordinates.
In \reffig[b,d]{fig:3}, in addition to lines with $\pm1$ slope extra features are observed for both $p,q < k$ (\reffig[b]{fig:3}) and $p,q > k$ (\reffig[d]{fig:3}).
Qualitatively these features can be labeled based on the knowledge of the mode frequencies in this driven configuration (colored arrows superimposed on \reffig[b,d]{fig:3} are in one-to-one correspondence with arrows in \reffig[a]{fig:3}).
 
By restricting and linearizing the full interaction Hamiltonian~(\ref{eq:hamiltonian-full}) to include only nearest-neighbor mode couplings -- e.g., $k$ with $k\pm\delta$, $k\pm\delta$ with $k\pm2\delta$, and so on -- we derive the analytical expression for the transmission spectrum (see discussion below \refeq{eq:hamiltonian} and \SM).
\reffig[c]{fig:3} shows the numerically calculated transmission spectrum, closely matching the experimental results in \reffig[d]{fig:3} and revealing non-Lorentzian lineshapes with multiple peaks.
The agreement between theory and experiment provides strong evidence that cascaded photon decays are the dominant mechanism underlying the non-equilibrium response.

\begin{figure*}
	\centering
	\includegraphics[width=4.76in]{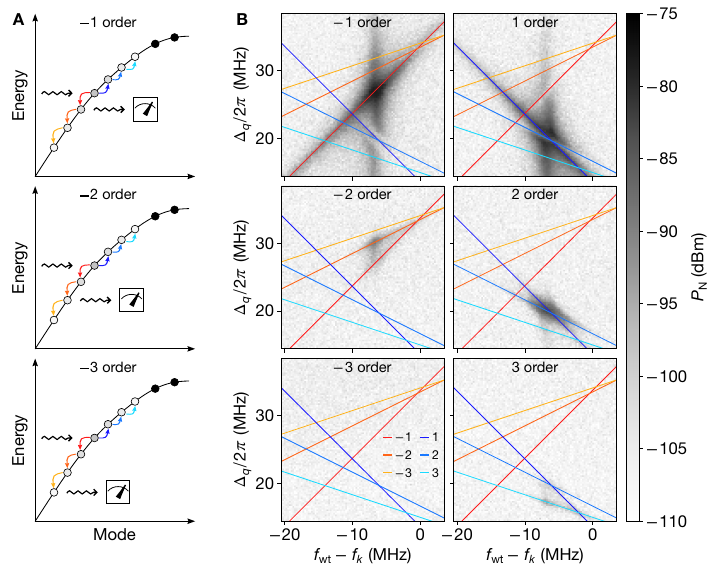}
	\caption{%
	\textbf{Direct observation of scattered photons.}  
	\textbf{(A)}~Schematic of the noise measurement $P_\mathrm{N}$ for cascaded scattering with orders $-1$, $-2$, and $-3$.
	Positive orders are measured analogously.
	\textbf{(B)}~Noise power $P_\mathrm{N}$ is measured for the processes with $|i| \leq 3$.
	The intensity of the color represents the number of photons emitted from the cavity, on top of a background consisting of amplified input-referred added noise (of $\approx -103$~dBm).
	The data in each panel are measured in the same pumping configuration, with the weak tone around $k=46$ and two pump tones $p=63$ and $q=66$. 
	Colored guides (same colors as in~\textbf{(A)}) show when the photon from the weak tone can be resonantly scattered to the mode $k + i \cdot \delta$.%
	}
	\label{fig:4}
\end{figure*}

This interpretation is further confirmed by directly detecting cascades to both lower and higher energy levels, as illustrated in \reffig[a]{fig:4}.  
Here, the noise power $P_\mathrm{N}$ emitted from the device serves as a measure of mode occupation.
Since the process of interest conserves energy and three tones are applied to the device input, the energy of the down- or up-scattered photon, $h\fany{\mathrm{N}}$, corresponding to order $i$, is unambiguously determined.  
Thus, $P_\mathrm{N}$ is measured in a narrow window of 270~Hz around the center frequency $\fany{\mathrm{N}}=\fany{\mathrm{wt}}+i\cdot\Delta/2\pi$, which is defined based on the pump tones and the frequency of the weak tone, $\fany{\mathrm{wt}}$.
To ensure the signal is easily resolvable, the power of the weak tone is slightly increased compared to the $S_{21}$ data presented earlier, resulting in a maximum mode $k$ occupation of approximately $1-2$~photons.
It was verified that, at this power level of the weak tone, $S_{21}$ remains within the linear response regime.

In \reffig[b]{fig:4}, the noise power $P_\mathrm{N}$ is plotted for cascaded scattering from mode $k=46$ when driving modes $p=26$ and $q=29$.
Guidelines overlaid on the data indicate the regions where energy conservation is satisfied for scattering processes of a given order $i$.
To determine the offsets of these lines, the mode frequencies for this driven configuration were independently identified.
Features corresponding to $\pm1$ order processes are highly pronounced in the respective subplots.
For higher-order processes of $\pm2$ and $\pm3$, distinct features emerge when the weak tone frequency $\fany{\mathrm{wt}}$ coincides with the mode $k$ (see \reffigS{9} for $S_{21}$ data measured in the same pumping configuration).
This behavior is expected, as the occupation of mode $k$ is highest when $\fany{\mathrm{wt}}$ is on resonance with mode $k$, and the number of scattered photons is proportional to $n_k$.
Interestingly, for $i=\pm1$ orders, a decrease in $P_\mathrm{N}$ is observed along the vertical feature when cascaded scattering with $|i| > 1$ becomes resonant with it suggesting that higher-order resonant scattering can dominate over non-resonant processes toward ``nearest-neighbors''.
In such resonant regime, the measured noise power suggests that nearly all incoming photons are efficiently scattered, indicating near-complete, cascaded down-conversion.
This is further supported by our estimates of conversion efficiency (see \SM).
Analogous measurement with $k > p, q$ is shown in \reffigS{8}, theoretical calculation of the same quantity is presented in \reffigS{11}.

\subsection*{Photonic kinetics}

\begin{figure*}
	\centering
	\includegraphics[width=4.76in]{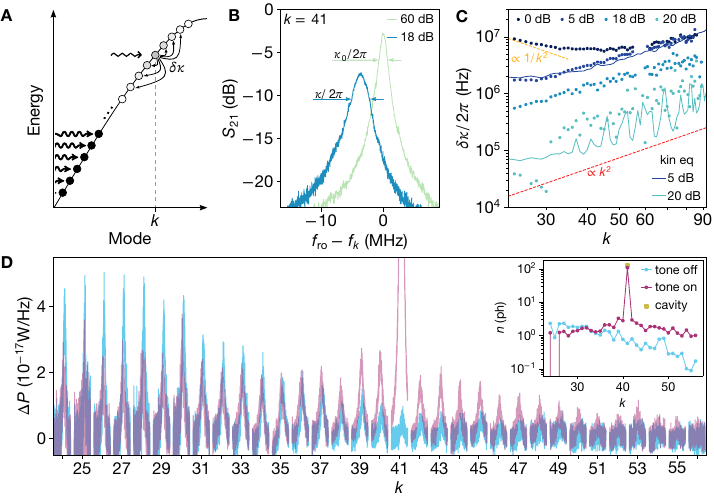}
	\caption{%
	\textbf{Incoherent broadband drive.}
	\textbf{(A)} Schematic of the experiment.
	The first 13 modes are pumped using an incoherent broadband source, with power controlled by a variable attenuator.
	Modes within the measurement band are probed with a weak tone, and the excess decay $\delta\kappa$ is extracted.
	\textbf{(B)} Under incoherent drive, the modes experience Kerr shift and an increase in linewidth.
	\textbf{(C)} Measured excess decay $\delta\kappa$ as a function of noise power (points).
	Solid lines represent results from simulations of kinetic equation, the dashed lines serve as a guide to the power law scaling of the decay.
	\textbf{(D)} Measurements of nonequilibrium occupations at 5~dB attenuation drive.
	The excess noise power spectral density, $\Delta P$, is measured with and without a weak tone applied to mode~41.
	Inset shows the nonequilibrium steady state distribution obtained by integrating the noise power spectral density from \textbf{(D)}.
	Yellow marker indicates occupation of mode~41 calculated using cavity coupling parameters and the weak tone power.%
	}
	\label{fig:5}
\end{figure*}

We have demonstrated that even two mode pumping is capable of enhancing the interaction part of the Hamiltonian~(\ref{eq:hamiltonian-full}), leading to the non-Lorentzian lineshapes from the cascaded redistribution of photons. 
Now, we turn to driving many modes, and studying the resulting near-continuum broadening of lineshapes.
This broadening emerges from intrinsic interaction-induced multimode scattering and is analogous to the quasiparticle lifetime of plasmons in one-dimensional quantum liquids~\cite{lin_prl_2013,bard_prb_2018} at equilibrium.
Critically, in our experiment the interplay of strong pumping, coupling of modes to waveguides, and non-trivial intrinsic relaxation due to multimode interactions drive the system far from equilibrium into a nonequilibrium steady state (NESS).
In the simplest approximation, such NESS can be described by the occupation numbers of modes from \refeq{eq:hamiltonian-full}, $n_k = \langle \adag{k} \anni{k}\rangle$, that is expected to be far from Bose-distribution function at the base temperature.
Below we will establish experimental ways of probing such steady state via broadening of lineshapes, mode occupation numbers and its redistribution in response to additional pumping.

To induce NESS and probe the continuous broadening of modes, we applied variable power broadband drive to a subset of modes  $k= 1,\ldots,13$ (\reffig[a]{fig:5}).
In this setup, a series of three amplifiers, acting as an incoherent drive source, is connected to the JJ-chain through a variable attenuator that controls the power of radiation entering the cryostat (see \reffigS{1}).
The maximum nominal attenuation (minimal power) of 60~dB gives the noise temperature of the radiation of approximately 300~K before entering the cryostat, with no observable change in mode frequencies or linewidths compared to undriven case.
For attenuations smaller than 30~dB, the modes exhibit a resolvable Kerr shift and at lower attenuations (higher noise power), an increase in linewidth is detected (\reffig[b]{fig:5}). 
The linewidth change is quantified via excess decay $\delta \kappa = \kappa - \kappa_0$, where $\kappa_0$ represents the linewidth measured at 60~dB attenuation.

In \reffig[c]{fig:5}, the excess decay, $\delta \kappa$, is plotted for different read-out modes $k$ and several attenuations. 
For weak drive powers (high attenuations), the broadening increases with wavenumber $k$, consistent with the interaction terms in \refeq{eq:hamiltonian-full} as coming from gradient nonlinearities~\cite{bard_prb_2018}.
Interestingly, the dependence on wavenumber is relatively weak, $\delta \kappa \propto k^2$, which contrasts with the theoretically predicted fourth-power scaling in the continuum regime~\cite{bard_prb_2018}.
To understand this excess decay quantitatively, we use kinetic equations~\cite{bard_prb_2018} to obtain the nonthermal distribution function in the NESS, $n_k$, and calculate the excess decay from the linearization of collision integral around the $n_k$. 
First, in \reffig[c]{fig:5} we match the observed scalings of excess decay for attenuation of 20 dB, using  a single fitting parameter that adjusts the occupation of the pumped modes (see \SM). 
At such low noise powers the intrinsic scattering is relatively weak, so excess decay is not a smooth function of the mode number, both, in theory and experiment. 
For intermediate-strength drives (e.g., 5~dB data), achieving similar match between experiment and theory requires the introduction of additional internal loss for the pumped modes of $\delta\ki=2\pi\!\times\!6$~MHz.
In the intermediate driving regime the excess decay can exceed the bare linewidth, and excess loss becomes a smooth function of the mode number, $\delta \kappa \propto k^2$, signaling that intrinsic scattering from interactions is strongly enhanced.
Finally, at the strongest drives in~\reffig[c]{fig:5} we observe a non-monotonic dependence of the excess decay $\delta \kappa$ on mode number.
Such qualitative change of excess decay cannot be explained by changes in NESS and signals presence of additional losses that are stronger for modes with low numbers, as anticipated with the introduction of $\delta \ki$ at intermediate driving strength.
We speculate that the excess decay may originate from interaction terms beyond the gradient nonlinearities included in \refeq{eq:hamiltonian-full}, such as phase slips.
Quantitatively, support for this hypothesis comes from the low-frequency power law, $\delta \kappa \propto  k^{-2}$ shown in \reffig[C]{fig:5}, which is consistent with the expected scaling due to phase slips \cite{bard_prb_2018}.
An appealing possible picture is that the effective chain parameters are renormalized out of equilibrium, leading to a larger effective phase slip rate compared to the near-equilibrium case.

In addition to the excess decay, we also characterize the strongly driven regime by the occupation of modes in the NESS using noise measurements.
We use the power spectral density (PSD) $P$ measured within a 15~MHz band around modes with numbers $k=24,\ldots, 56$, both with the drive applied (5~dB attenuation) and without the drive (60~dB)~\cite{vrajitoarea_arxiv_2022,fraudet_prl_2025}.
The difference, $\Delta P = P(5~\mathrm{dB}) - P(60~\mathrm{dB})$, is plotted in \reffig[D]{fig:5} (``tone off'').
Crucially, this data can be used to estimate the excess occupation on top of thermal distribution in the NESS.
Ignoring thermal occupation at large mode numbers we obtain the estimate of occupation numbers $n_k \approx \kex^{-1} \int df\, \Delta P / (G hf) $, where the integration is performed over a window 15~MHz around $f'_k$, and $G$ is the input-referred calibrated gain of the amplifier (\reffigS{10}).
This estimation has a systematic error of approximately a few~dB, and a more sophisticated method should be employed to account for the losses of cryogenic components before the amplifier~\cite{brubaker_prl_2017,leonard_arxiv_2024}. 

The occupation of mode numbers shown in~\reffig[D]{fig:5} reveals at least one order  of magnitude increase in $n_k$ compared to thermal distribution, even for modes with numbers $k=30$ that are far from pumped modes.
Experimental imperfections, such as leakage outside the passband of the low-pass filter, cannot explain this excess occupation.
However the large enhancement of $n_k$ for $k$ far away from pumped modes is consistent with intrinsic redistribution of photons due to interaction-induced scattering.
Intuitively, photons injected by driving have enough opportunities to participate in multiple two photon-scattering events encoded by the interaction term in \refeq{eq:hamiltonian-full}, that redistribute excitations from modes $m, n\to k,l$.
This causes increase in the occupation of modes with $k\gg 13$ observed experimentally in~\reffig[D]{fig:5}.
The redistribution of excitations due to enhanced intrinsic scattering is also revealed by the response of NESS to additional tone applied to mode $k=41$.
The ``tone on'' data in \reffig[D]{fig:5} shows the global change in PSD as a response to additional driving of mode $41$.
Converting this data to occupation numbers in the NESS, we observe that $n_k$ changes for all modes within the spectral range, witnessing non-local redistribution of excitations due to multiple interaction-induced scattering. 
Also, additional driving of this mode verified the occupation estimates: the independently calculated occupation of mode~41 from the calibrated insertion loss (yellow marker in the inset of \reffig[d]{fig:5}) agrees well with the PSD measurement.

\section*{Discussion}
Summarizing, we have established that the quantized plasma modes in the Josephson junction chain as a practical realization of a driven, multimode nonlinear system.
Using multimode microwave spectroscopy techniques, we characterized the Hamiltonian of the system in the regime of moderate interactions, directly observing cascaded couplings of plasma modes, and experimentally identifying the microscopic origin of interaction term.
Subjecting the system to strong drives, we created non-equilibrium steady state, where drastic enhancement of intrinsic interactions becomes a limiting factor of plasma mode lifetimes.
Direct visualization of nonthermal mode occupations and resulting smooth mode broadenings reveals realization of non-equilibrium strongly interacting liquid in a multimode superconducting resonator. 

Although our system is quantum, it is important that many of our observables can be understood classically, and do not rely on entanglement or negativity.
The few-mode driving regime is described by linearized equations of motion, which can be viewed as classical equations for wave amplitudes.
The steady state for many-mode drives is described by a kinetic equations for plasmon occupation number which is similar to classical kinetics for wave action \cite{nazarenko_book_2011}, albeit with some modifications due to quantum effects.
We view the current work as a necessary starting point for future investigations of ``more quantum" observables and regimes, such as intermode coherences, regime of small occupation numbers, and transient dynamics.
The transient dynamics in particular can be be strongly nonclassical even for weak nonlinearities \cite{lingenfelter_unconditional_2021}.

Looking further ahead, our work opens a number of exciting new directions. 
We demonstrated that driving causes the broadening from momentum-conserving interactions $\delta \kappa$ to exceed all other sources, suggesting realization of hydrodynamic regime for plasmons. 
This invites detailed characterization of non-equilibrium steady state and search for emergence of the universal scaling regime as expected from the classical wave turbulence~\cite{nazarenko_book_2011}. 
Moreover, future work can explore the vast space of coherent and incoherent driving configurations, characterizing crossover between regimes of weak and strong multimode scattering, as well as potential transitions between different universal regimes.

In a different direction, coherent targeted driving of small subset of modes may allow to implement squeezing, and creation of other nonclassical states and their characterization.
Also, realization of non-reciprocal interactions between modes may lead to non-Hermitian lattice models~\cite{mcdonald_natcom_2020}. 
Finally, mixing coherent and incoherent driving may enable studies of engineered wave turbulence with non-reciprocal interactions~\cite{dewit_nature_2024}, and improve our understanding of microscopic dissipation mechanism in superconducting resonators.

More broadly, our work expands the experimental flexibility of Josephson arrays known for realizing different models~\cite{bell_comprenphys_2018,chamon_prxquantum_2021,chandra_prl_1995} by adding powerful capabilities to induce crossover from weakly interacting multimode system to strongly coupled one-dimensional liquid.
Combined with the flexible experimental toolbox of microwave control with circuit quantum electrodynamics~\cite{blais_rmp_2021}, this expansion sets the stage for an exciting new chapter of experimental exploration of strongly interacting quantum liquids.

\onecolumngrid
\section*{Materials and Methods}
\paragraph*{Device fabrication:}
The device was fabricated on a highly resistive silicon substrate using standard nanofabrication techniques.
E-beam lithography (Raith EBPG5150) was performed on a resist stack consisting of LOR 5B/W/CSAR~62 with thicknesses of 440/40/100~nm, respectively~\cite{negrov_rusmicro_2016}.
The high-contrast CSAR~62 layer was used to define the pattern on the W layer, which was subsequently etched using reactive ion etching with inductively coupled plasma (RIE-ICP) in SF$_6$.
In the exposed regions the LOR~5B was removed with O$_2$ ICP, followed by a short dip in diluted MF319 (MF319:H$_2$O = 1:2) to eliminate any residual LOR 5B.
After this step the undercut was created and suspended W bridges are formed.
Double-angle Al deposition (25/50~nm at $\pm23^\circ$) was carried out in a UHV e-beam evaporator (Plassys MEB550S2).
After the first Al deposition static oxidation was performed at 5~mbar for 5~minutes.
Lift-off was completed in successive hot baths of DMSO, acetone, and IPA.

\paragraph*{Microwave spectroscopy:}
Microwave measurements were carried out in an Oxford Triton dilution refrigerator with a base temperature of approximately 12~mK.
The microwave wiring of the refrigerator is shown in \reffigS{1}.
For three-tone spectroscopy (\reffig{fig:1}, \reffig{fig:2}, \reffig{fig:3}), signals from two signal generators and a vector network analyzer were combined using a power splitter and a directional coupler.
For noise measurements (\reffig{fig:4}), the output signal was directed to a spectrum analyzer while the rest of the measurement chain remained identical to that used in Figs.~1-3.
Broadband pumping was implemented using a cascade of room-temperature amplifiers, with the total power entering the cryostat controlled by a variable attenuator.
The insertion loss of the cryostat was measured using throughline calibration and together with the loss of room temperature components was used to infer device input referred power.

\paragraph*{Theory:}
We model the Josephson chain plasmons, interacting via quartic order nonlinearities, through the Hamiltonian in \refeq{eq:hamiltonian-full},
\begin{equation}
H = \sum_{k>0} \hbar \omega_k \adag{k} \anni{k} +  \sum_{klmn} \hbar K_{klmn} \adag{k} \adag{l} \anni{m} \anni{n}.
\end{equation}
The matrix element describing $2\to2$ scattering is given by
\begin{equation}
 K_{klmn}=-\frac{\Eg\pi^2}{4N^3} \sqrt{klmn}\sum_{s_1,s_2,s_3=\pm}\delta_{k+s_1l+s_2m+s_3n,0},
\end{equation}
where $k,l,m,n$ are mode numbers, $\Eg$ is the ground charging energy and $N$ the number of junctions.

To reproduce the weak drive kinetics, we use quantum Langevin equations and the input-output formalism, treating the two pumped modes as classical oscillating fields. Setting $i_\text{max}$ to be the maximum order of included multi-mode interaction, the transmission amplitude for the readout mode $k$ is
\begin{equation}
    S_{21}[\omega]=\mathrm{i}S_\text{fit} \kappa_{\text{ex},k}\left[(\omega\mathbb{I}-\mathbb{A}_\text{fit})^{-1}\right]_{i_\text{max}+1,i_\text{max}+1}e^{\mathrm{i}\varphi_\text{fit}},
\end{equation}
with
\begin{equation}
\begin{cases}
    &\mathbb{A}_\text{fit}[n,n]=\omega_{k+(n-1-i_\text{max})\delta}(1+\alpha_\text{fit})-(n-1-i_\text{max})\Delta-\mathrm{i}\frac{\kappa_\text{fit}}{2}\\
    &\mathbb{A}_\text{fit}[m,m-1]=\mathbb{A}[m-1,m]=K_{p,p+\delta,k+(m-1-i_\text{max})\delta,k+(m-2-i_\text{max})\delta}n^\text{pump}_\text{fit}.
\end{cases}
\end{equation}
The matrix indices $n$ and $m$ run in the range $n\in[1,2i_\text{max}+1],\,m\in[2,2i_\text{max}+1]$. $\Delta$ is the difference between the pump frequencies, $\delta$ the distance between the pumped modes, and the five fitting parameters are emphasized by the corresponding subscript.

Moving to the strong multi-mode drive regime, we calculate the excess linewidth due to four-wave mixing -- excluding self-decays -- as
\begin{equation}
\begin{aligned}
\delta\kappa_k
&= \frac{\pi^5 \Eg^2}{16N^6}
   \sum_{p,q_1,q_2} kpq_1q_2 \,
   \delta(\omega_k+\omega_p-\omega_{q_1}-\omega_{q_2})
   \bigl[n_p(1+n_{q_1}+n_{q_2})-n_{q_1}n_{q_2}\bigr]
   \sum_{s_1,s_2,s_3=\pm}
   \delta_{k+s_1p+s_2 q_1+s_3 q_2,0}.
\end{aligned}
\end{equation}
The occupation numbers correspond to the nonequilibrium steady state of the system. This is obtained numerically by solving the kinetic equation
\begin{equation}
    \dot n_k = - (\kappa_k+\delta\kappa_{\mathrm{i},k}) (n_k-n_k^\text{th}) 
    + \alpha\kappa_{\text{ex},k}  n^\text{flux}_k
    + I_\text{tot}[k],
\end{equation}
with $n^\text{th}$ the Bose-Einstein occupation, $n^\text{flux}$ the injected photons per unit frequency and time, $\kappa$ the bare linewidth and $\kex$ the coupling rate to each terminal. $I_\text{tot}[k]$ is the total collision integral
\begin{equation}
I_\text{tot}[k] 
=
\sum_{p,q_1,q_2} W_{ p,k \to q_1q_2} \{(1+n_p)n_{q_1} n_{q_2}-n_k[n_p(1+n_{q_1}+n_{q_2})-n_{q_1}n_{q_2}]\},
\end{equation}
where the transition probability $W$ is based on the above matrix element $K$. The conservation of energy during collisions is modeled as a Lorentzian with linewidth given by the sum of the linewidths of the four participating modes,
\begin{equation}
	\delta_\gamma(\omega)= \frac{1}{\pi}\frac{\gamma}{\gamma^2+\omega^2},\qquad\gamma=\sum_{j\in[k,p,q_1,q_2]}(\kappa_j+\delta\kappa_j)/2,\qquad\omega=\omega_{q_1}+\omega_{q_2}-\omega_k-\omega_p.
\end{equation}
Importantly, such linewidths are updated every ten time-steps of the evolution to include the excess linewidth $\delta \kappa_j$ due to the intrinsic nonlinear scattering. Lastly, the parameter $\alpha=5$ is a frequency-independent correction to the insertion loss of the fridge, determined by comparing the 20~dB data and simulations. For stronger drives (5~dB), we introduce an additional internal loss $\delta\kappa_{\mathrm{i},k}=6$~MHz for the driven modes ($k=1,\ldots,19$).
\twocolumngrid

\bibliography{main.bib} 

\section*{Acknowledgments}
\paragraph*{Discussions:} We thank Vincenzo Vitelli, Michel Fruchart, and Amir Burshstein for helpful input.
\paragraph*{Technical support:} We acknowledge technical support from the Nanofabrication Facility and the MIBA machine shop at IST Austria.
\paragraph*{Funding:} This research was supported in part by grant NSF PHY-2309135 to the Kavli Institute for Theoretical Physics (KITP), by the Austrian Science Fund (FWF) SFB F86, and by the NOMIS foundation.
\paragraph*{Author Contributions:}
Writing--original draft: AVB, LV, MS, APH;
Conceptualization: AVB, LV, MS, APH;
Investigation: AVB, LV, MS, APH;
Writing--review \& editing: AVB, LV, MS, APH;
Methodology: AVB, LV, MS, APH;
Resources: AVB, MS, APH;
Funding acquisition: MS, APH;
Data curation: AVB, MS, APH;
Validation: AVB, LV, MS;
Supervision: MS, APH;
Formal analysis: AVB, LV, MS;
Software: AVB, LV, MS, APH;
Project administration: APH;
Visualization: AVB, LV, APH.

\paragraph*{Competing interests:} There are no competing interests to declare.
\paragraph*{Data and materials availability:} Data are available at doi:10.5281/zenodo.17009557.
For the purpose of open access, the authors have applied a CC BY public copyright license to any Author Accepted Manuscript version arising from this submission.

\section*{Supplementary Materials}
\newcommand{\suppinclude}{
\noindent
Supplementary Text\\
Figures S1 to S11\\
Table S1\\
References \textit{(65-68)}}
\suppinclude

\clearpage
\onecolumngrid
\renewcommand{\thefigure}{S\arabic{figure}}
\renewcommand{\thetable}{S\arabic{table}}
\renewcommand{\theequation}{S\arabic{equation}}
\renewcommand{\thepage}{S\arabic{page}}
\setcounter{figure}{0}
\setcounter{table}{0}
\setcounter{equation}{0}
\setcounter{page}{1}

\def\scititle{
	Nonequilibrium plasmon liquid in a Josephson junction chain
}

\begin{center}
\section*{Supplementary Materials for\\ \scititle}
Anton V. Bubis,
Lucia Vigliotti,
Maksym Serbyn,
Andrew P. Higginbotham$^\ast$\\
\small$^\ast$Corresponding author. Email: ahigginbotham@uchicago.edu
\end{center}

\subsection*{This PDF file includes:}
\suppinclude
\newpage


\section*{JJ-chain parameters}
\label{sec:jjchain_params}
First 173 modes were identified using the two-tone spectroscopy technique (see, for example, the Supplementary Information of Ref.~\cite{kuzmin_natphys_2019}).
To obtain frequencies unaffected by the Kerr shift due to pump, the pump power must be sufficiently low.
Thus, both the pump power (1~dB step) and pump frequency (1~MHz step) were swept while recording $S_{21}$ at a fixed $\fro$ (on resonance with the undriven mode 30).
For each mode observed in the two-tone data, the lowest pump power was selected where a dip in the phase of $S_{21}$ was detected above the noise floor. 
Mode frequencies obtained by this approach are plotted in (\reffig{fig:dispersion}).
The dispersion of the chain in the limit of $k \ll N$, where $N$ is the number of junctions (for our chain $N = 13157$), is well-known~\cite{masluk_prl_2012}:
\begin{equation}
	2\pi \fany{k} = \frac{v\mathbf{k}}{\sqrt{1 + \left(\frac{v\mathbf{k}}{\omega_\mathrm{p}}\right)^2}},\quad\mathbf{k} = k \frac{\pi}{L}.
	\label{eq:dispersion}
\end{equation}
By fitting to \refeq{eq:dispersion}, the speed of light $v$ and the plasma frequency of a single junction $\omega_\mathrm{p}$ were extracted.

A complete description of the JJ-chain requires three independent energy scales: the Josephson energy $\Ej$, the junction charging energy $\Ec$, and the ground charging energy $\Eg$~\cite{masluk_prl_2012}.
However, $v$ and $\omega_\mathrm{p}$ depend on the products $\Ej\Eg$ and $\Ej\Ec$, meaning that the chain parameters cannot be independently extracted.
Additional measurements are necessary to determine relevant energies independently, for example, through DC measurements of the chain (see Supplementary Information in Ref.~\cite{mukhopadhyay_natphys_2023}).

To fix device parameters we calculate $E_g$ based on the device geometry, and use the fit to determine the remaining unknown parameters $E_J$ and $E_C$.
$E_g$ is found from standard transmission line formulas~\cite{gevorgian_ieee-microwave_1995,simons_book_2001}.
To check the result, we also simulated our device geometry in Sonnet, finding a value that agrees with the analytical formula to within 1\%.
The standard error of the fit plasmon speed $v$ and the plasma frequency $\omega_\mathrm{p}$ are also less than 1\%, making it tempting to infer an overall parameter error of only a few percent.
However, caution in interpreting the standard errors is needed because the measured dispersion (\reffig[b]{fig:dispersion}) shows a systematic discrepancy with the theory.

As a rough consistency check, we also estimate the capacitance of a single junction from its geometry together with a nominal specific capacitance of Al/AlO$_\mathrm{x}$/Al junctions, using the typical range 50-100~fF/\textmu m$^2$~\cite{kerr_ieee-microwave_1992,deppe_determination_2004}, determining the remaining two parameters $E_g$ and $E_J$ from the fit.
It is comforting that the rough consistency check overlaps with the more precisely-determined value based on device geometry.
Both approaches follow the methodology outlined in Ref.~\cite{kuzmin_natphys_2019}.

\section*{Derivation of the effective Hamiltonian and matrix element}
\label{sec:Matrix_element}
In this Section, we derive the matrix element $K$ responsible for the beam-splitter-like interaction. Our starting point is the following Luttinger liquid quadratic Hamiltonian, describing the JJ chain in the continuum~\cite{bard_prb_2018} and in the absence of nonlinearities,
\begin{equation}
    H_0 =\frac{1}{2\pi^2} \int_{0}^{N} dx \left[\frac{v_s}{K_g} |\partial_x\phi(x)|^2 +v_sK_g\pi^2|\partial_x\theta(x)|^2\right].\label{eqn:H0}
\end{equation}
Here $K_\mathrm{g}=\sqrt{{\Ej}/({2\Eg})}$, $v_s=\sqrt{2\Ej \Eg}$, and $N$ is the number of Josephson junctions. $\theta(x)$ is the superconducting phase field, and $\phi(x)$, related to the number of Cooper pairs, is its canonically conjugated. We set $\hbar=1$, and express distances in units of the junctions' spacing $L/N$, where $L$ is the chain's size. With this choice, $ x= r\frac{N}{L}$, where $r$ has dimension of length. The fields $\phi$ and $\theta$ can be expanded on bosonic operators describing the plasmonic modes sustained by the chain,
\begin{align}
    \theta(x)&=\mathrm{i}\left(\frac{2\Eg}{\Ej}\right)^{1/4}\frac{1}{\sqrt{N}}\sum_{\textbf{k}>0} \frac{1}{\sqrt{\textbf{k}}}\sin{(\textbf{k}x)} (\hat{a}_{\textbf{k}}^{\dagger} - \hat{a}_{\textbf{k}}),\label{eqn:theta}\\
    \phi(x) &= \pi\left(\frac{\Ej}{2\Eg}\right)^{1/4} \frac{1}{\sqrt{N}}\sum_{\textbf{k}>0}\frac{1}{\sqrt{\textbf{k}}}\cos{(\textbf{k}x)} (\hat{a}_{\textbf{k}}^{\dagger} + \hat{a}_{\textbf{k}}).\label{eqn:phi}
    \end{align}
$\textbf{k}$ is the quasimomentum, which can be quantized as $\textbf{k}=\pi k/N$, where $k=1,2,\dots$ are the corresponding mode numbers. The above expansions rely on the plasmonic wavefunctions under open boundary conditions, which take the following form
\begin{equation}
    \psi_k(x)=\sqrt{\frac{2}{N}}\sin{\left(\frac{\pi kx}{N}\right)}.
\end{equation}
After substitution of Eqs.~(\ref{eqn:theta},\ref{eqn:phi}) in \refeq{eqn:H0}, we get the second quantized Hamiltonian
    \begin{equation}
        H_0=\sum_{{k}>0}v_s {k}\left(\hat{a}_{{k}}^{\dagger}\hat{a}_{{k}}+\frac{1}{2}\right)= \sum_{k>0}  \underbrace{\sqrt{2\Eg \Ej}\frac{\pi}{N}k}_{\omega_k} \left(\hat{a}^\dagger_k \hat{a}_{k} +\frac12\right),\label{eqn:h0_2ndq}
    \end{equation}
which allows us to identify mode frequencies in our notations, $\omega_k\approx v_sk$. Notice that \refeq{eqn:H0} was already assuming a linearized plasmonic dispersion. If we also apply second quantization to the leading \emph{nonlinear} term from the expansion of the cosine Josephson potential,
\begin{equation}
 H_\text{int} = - \frac{\Ej}{24}\int_0^N dx \left(\partial_x\theta\right)^4,
\end{equation}
we get
    \begin{equation}
        H_\text{int}=- \frac{\Eg\pi^2}{96N^3}\sum_{k,l,m,n>0}\sqrt{klmn}(\hat{a}_{k}^{\dagger} - \hat{a}_{k})(\hat{a}_{l}^{\dagger} - \hat{a}_{l})(\hat{a}_{m}^{\dagger} - \hat{a}_{m})(\hat{a}_{n}^{\dagger} - \hat{a}_{n})\sum_{s_1,s_2,s_3=\pm}\delta_{k+s_1l+s_2m+s_3n,0}.\label{eqn:nl_2ndq}
    \end{equation}
We have verified that $1\to 3$ scattering -- arising from $\hat{a}_k\hat{a}_l\hat{a}_m\hat{a}_n^{\dagger}$ -- is energetically much less favorable than $2\to 2$, though arising at the same nonlinear order. For this reason we neglect it in our analysis. The four-wave mixing matrix element is
     \begin{equation}
    K_{klmn}=\langle 0| \hat{a}_k \hat{a}_l\hat{H}_\text{int} \hat{a}_m^\dagger \hat{a}_n^\dagger |0\rangle =-\frac{\Eg\pi^2}{4N^3} \sqrt{klmn}\sum_{s_1,s_2,s_3=\pm}\delta_{k+s_1l+s_2m+s_3n,0}.\label{eqn:boxmatrixel}
    \end{equation}
Notice that, for plasma modes in a confined system, quasimomentum is conserved only modulo sign, as each standing wave supports two components propagating in opposite directions. The expression for $K_{klmn}$ determines the nonlinear term in the full Hamiltonian in \refeqS{1} in the main text (obtained after restriction to boson-number conserving terms only), and its restriction to the momentum and energy conserving processes results in \refeqS{2} and will be used below to calculate the analytical predictions for the transmission.

We have verified that the matrix element associated with sixth order nonlinearities is strongly suppressed compared to that of the fourth order. This justifies restricting our analysis to the fourth order interaction. Furthermore, when investigating the excess linewidths (Sec. SM V), we checked that the extra broadening due to sixth order processes is negligible.

\section*{Quantum Langevin equations}
\label{sec:Langevin}

Let us start from the Hamiltonian describing JJ-chain expanded up to the fourth order (see~\refeqS{2}):
\begin{align}
	H = H_0+H_\mathrm{int}= \sum_{k>0} \hbar \omega_k \adag{k} \anni{k} +  \sum_{klmn} \hbar K_{klmn} \adag{k} \adag{l} \anni{m} \anni{n},\label{eq:hamiltonian_init}
\end{align}
where $H_\mathrm{int}$ is a nonlinear contribution originated from the expansion of the Josephson cosine potential, and it contains all permutations of four bosonic operators (quartic terms).
To illustrate the four-wave mixing process of the Hamiltonian \refeq{eq:hamiltonian_init}, let us focus on the process $i,p \leftrightarrow j,q$:
\begin{align*}
	H = \sum_{k=i,j,p,q} \hbar \omeg{k} \adag{k}\anni{k} + \hbar K_{ijpq} \left(\adag{i} \anni{j} \adag{p} \anni{q} + \anni{i} \adag{j} \anni{p} \adag{q}\right),
\end{align*}
where we leave just four modes of interest and assume that mode frequencies $\omeg{k}$ already include Kerr shifts (highlighted with prime index in the main text).
For both pumps of modes $p$ and $q$ we can use classical fields $A_p e^{-\mathrm{i}\omeg[\mathrm{pump}]{p}t}$ and $A_q e^{-\mathrm{i}\omeg[\mathrm{pump}]{q}t}$, respectively.
After this substitution, the Hamiltonian reads:
\begin{align*}
	H = \hbar \left(\omeg{i} \adag{i} \anni{i} + \omeg{j} \adag{j} \anni{j} + g \adag{i} \anni{j} e^{-\mathrm{i}\Delta t} + g^{*} \anni{i} \adag{j} e^{\mathrm{i}\Delta t}\right),
\end{align*}
where $g=K_{ijpq} A_{p}^{*} A_q$, $\Delta=\omeg[\mathrm{pump}]{q}-\omeg[\mathrm{pump}]{p}$.
Now we need to solve quantum Langevin equations for modes $k=i, j$:
\begin{align*}
	\frac{d\anni{k}}{dt} = -\frac{\mathrm{i}}{\hbar} \left[\anni{k}, H \right] - \frac{\kappa_{\mathrm{L},k}+\kappa_{\mathrm{R},k} + \kappa_{\mathrm{i},k}}{2} \anni{k} + \sqrt{\kappa_{\mathrm{L},k}}\anni{\mathrm{IN, L}} + \sqrt{\kappa_{\mathrm{R},k}}\anni{\mathrm{IN, R}} + \sqrt{\kappa_{\mathrm{i},k}}\fnoise{\mathrm{IN}}.
\end{align*}
Here $\anni{\mathrm{IN, L}}, \anni{\mathrm{IN, R}}$ are input fields on the left and right side of the cavity, and $\fnoise{\mathrm{IN}}$ is an effective Langevin force which describes all loss mechanisms.
Assuming for simplicity mode independent linewidth $\kappa_{(\mathrm{R,L}),k} = \kex$, $\kappa_{\mathrm{i},k} = \ki$  and $\kappa = 2\kex + \ki$ we obtain:
\begin{align}
	\frac{d\anni{i}}{dt} &= -\mathrm{i}\omeg{i}\anni{i} - \mathrm{i}g\anni{j}e^{-\mathrm{i}\Delta t} - \frac{\kappa}{2}\anni{i} + \sqrt{\kex}\anni{\mathrm{IN, L}} + \sqrt{\kex}\anni{\mathrm{IN, R}} + \sqrt{\ki}\fnoise{\mathrm{IN}},\label{eq:Langevin1}\\
	\frac{d\anni{j}}{dt} &= -\mathrm{i}\omeg{j}\anni{j} - \mathrm{i}g^{*}\anni{i}e^{\mathrm{i}\Delta t} - \frac{\kappa}{2}\anni{j} + \sqrt{\kex}\anni{\mathrm{IN, L}} + \sqrt{\kex}\anni{\mathrm{IN, R}} + \sqrt{\ki}\fnoise{\mathrm{IN}}.\label{eq:Langevin2}
\end{align}
After taking Fourier transform of Eqs.~(\ref{eq:Langevin1})-(\ref{eq:Langevin2}):
\begin{align*}
	-\mathrm{i} \omega \anni{i}[\omega] &= -\mathrm{i}\omeg{i}\anni{i}[\omega] - \mathrm{i}g\anni{j}[\omega - \Delta]- \frac{\kappa}{2}\anni{i}[\omega] + \sqrt{\kex}\anni{\mathrm{IN, L}}[\omega] + \sqrt{\kex}\anni{\mathrm{IN, R}}[\omega] + \sqrt{\ki}\fnoise{\mathrm{IN}}[\omega],\\
	-\mathrm{i} \omega \anni{j}[\omega] &= -\mathrm{i}\omeg{j}\anni{j}[\omega] - \mathrm{i}g^{*}\anni{i}[\omega + \Delta] - \frac{\kappa}{2}\anni{j}[\omega] + \sqrt{\kex}\anni{\mathrm{IN, L}}[\omega] + \sqrt{\kex}\anni{\mathrm{IN, R}}[\omega] + \sqrt{\ki}\fnoise{\mathrm{IN}}[\omega].
\end{align*}
Finally, introducing notations:
\begin{align*}
	\chii[(3)]{i}[\omega] &= \frac{1}{\kappa/2 - \mathrm{i}(\omega - \omeg{j} - \Delta)},\\
	\chii[(3)]{j}[\omega] &= \frac{1}{\kappa/2 - \mathrm{i}(\omega - \omeg{i} + \Delta)},\\
	\chii{k}[\omega] &= \frac{1}{\kappa/2 - \mathrm{i}(\omega - \omeg{k}) + |g|^2\chii[(3)]{k}[\omega]},
\end{align*}
we obtain:
\begin{align*}
	\anni{i}[\omega] &= \chii{i}[\omega] \left[ \left(\sqrt{\kex}\anni{\mathrm{IN, L}} + \sqrt{\kex}\anni{\mathrm{IN, R}} + \sqrt{\ki}\fnoise{\mathrm{IN}} \right)[\omega] - \mathrm{i}g\chii[(3)]{i}[\omega]\left( \sqrt{\kex}\anni{\mathrm{IN, L}} + \sqrt{\kex}\anni{\mathrm{IN, R}} + \sqrt{\ki}\fnoise{\mathrm{IN}} \right)[\omega-\Delta]\right],\\
	\anni{j}[\omega] &= \chii{j}[\omega] \left[ \left(\sqrt{\kex}\anni{\mathrm{IN, L}} + \sqrt{\kex}\anni{\mathrm{IN, R}} + \sqrt{\ki}\fnoise{\mathrm{IN}} \right)[\omega] - \mathrm{i}g^*\chii[(3)]{j}[\omega]\left( \sqrt{\kex}\anni{\mathrm{IN, L}} + \sqrt{\kex}\anni{\mathrm{IN, R}} + \sqrt{\ki}\fnoise{\mathrm{IN}} \right)[\omega+\Delta]\right].
\end{align*}
Using boundary conditions:
\begin{align*}
	\anni{\mathrm{IN, R}}+ \anni{\mathrm{OUT, R}} = \sqrt{\kex}\anni{i} + \sqrt{\kex}\anni{j},\\
	\anni{\mathrm{IN, L}} + \anni{\mathrm{OUT, L}} = \sqrt{\kex}\anni{i} + \sqrt{\kex}\anni{j},
\end{align*}
we can express averaged output field $\langle \anni{\mathrm{OUT, L}} \rangle$ via input $\langle \anni{\mathrm{IN, R}} \rangle$:
\begin{align}
	S_{21}[\omega] = \frac{\langle \anni{\mathrm{OUT, L}} \rangle[\omega]}{\langle \anni{\mathrm{IN, R}} \rangle[\omega]} = \kex (\chii{i}[\omega]+\chii{j}[\omega]) \approx \frac{\kex}{\kappa/2 - \mathrm{i}(\omega-\omeg{i}) + \frac{|g|^2}{\kappa/2-\mathrm{i}(\omega-\omeg{j}-\Delta)}},\label{eq:transmission}
\end{align}
where in the last expression we ignored the small contribution from $\chii{j}[\omega]$, because we are interested in a solution in the vicinity of mode $i$.
This expression is used to extract matrix element from experimental data e.g. in \reffigS[D]{2} of the maintext.

\section*{Generalization of quantum Langevin equations to multiplets of modes}
\label{sec:langevin_cascades}

This Section is devoted to the theoretical description of the cascaded process shown in Fig. 3 of the main text. In this process, the excitation down or up-scatters from mode $k$ to mode $k\mp i\delta$, and a simultaneous up-scattering or down-scattering from mode $p$ to mode $p\pm \delta$ occurs $i$ times in order to satisfy momentum conservation. Schematically this process may be written as:
\begin{align} \label{Eq:process1}
    &k\to k-i\delta,\qquad [p\to p+\delta]^i\\
     \label{Eq:process2}
    &k\to k+i\delta,\qquad [p+\delta\to p]^i.
\end{align}
Here $p, q=p+\delta$ are the pumped modes, $k$ is the readout mode, and $i\in[1,i_\text{max}]$ is a positive integer. The upper bound $i_\text{max}$ sets to which order the multi-mode interaction is included. 

In order to account for processes in Eqs.~(\ref{Eq:process1})-(\ref{Eq:process2}), we can restrict the general Hamiltonian (\refeqS{1} in the main text) to the following set of modes
\begin{equation}\label{Eq:M-def}
\mathcal{M} = \{p, p+\delta, k-i_\text{max}\delta,k-(i_\text{max}-1)\delta,\ldots,k-\delta, k, k+\delta,\ldots,k+(i_\text{max}-1)\delta,k+i_\text{max}\delta\}.
\end{equation}
The Hamiltonian describing such set reads (we set $\hbar=1$)
\begin{multline}
H_\mathcal{M} =\sum_{j \in \mathcal{M}} \omega_j \hat{a}^\dagger_j \hat{a}_j+\\
A_p e^{-\mathrm{i}\omega^\mathrm{pump}_p t} A^*_{p+\delta} e^{\mathrm{i}\omega^\mathrm{pump}_{p+\delta} t}\sum_{i=1}^{i_\text{max}}
\big[K_{p,p+\delta,k+(i-1)\delta,k+i\delta} \hat{a}^\dagger_{k+(i-1)\delta} \hat{a}_{k+i\delta}
 +K_{p,p+\delta,k-(i-1)\delta,k-i\delta} \hat{a}^\dagger_{k-i\delta} \hat{a}_{k-(i-1)\delta}\big]+\text{h.c.},
\end{multline}
where we have replaced the two pumped modes by classical fields. The Langevin equations for all the other modes in $\mathcal{M}$ reads (we omit the Langevin force $\hat{f}_\text{IN}$ for brevity)
\begin{eqnarray*}
\dot{\hat{a}}_{k-i_\text{max}\delta} &=& -\mathrm{i}\omega_{k-i_\text{max}\delta} \hat{a}_{k-i_\text{max}\delta}- \mathrm{i} K_{p,p+\delta,k-i_\text{max}\delta,k-(i_\text{max}-1)\delta} A_p A^*_{p+\delta}e^{-\mathrm{i}(\omega^\mathrm{pump}_p-\omega^\mathrm{pump}_{p+\delta}) t} \hat{a}_{k-(i_\text{max}-1)\delta}-\frac{\kappa_{k-i_\text{max}\delta}}{2}\hat{a}_{k-i_\text{max}\delta},\\
 \dot {\hat{a}}_{k-i\delta} &=& -\mathrm{i}\omega_{k-i\delta} \hat{a}_{k-i\delta} - \mathrm{i} K_{p,p+\delta,k-i\delta,k-(i-1)\delta} A_p A^*_{p+\delta} e^{-\mathrm{i}(\omega^\mathrm{pump}_p-\omega^\mathrm{pump}_{p+\delta}) t}\hat{a}_{k-(i-1)\delta}\\
 &&- \mathrm{i} K_{p,p+\delta,k-(i+1)\delta,k-i\delta} A^*_p A_{p+\delta} e^{\mathrm{i}(\omega^\mathrm{pump}_p-\omega^\mathrm{pump}_{p+\delta}) t}\hat{a}_{k-(i+1)\delta}-\frac{\kappa_{k-i\delta}}{2}\hat{a}_{k-i\delta},\qquad i\neq i_\text{max}\\
 \dot {\hat{a}}_k &=& -\mathrm{i}\omega_k \hat{a}_k - \mathrm{i} K_{p,p+\delta,k,k+\delta} A_p A^*_{p+\delta}e^{-\mathrm{i}(\omega^\mathrm{pump}_p-\omega^\mathrm{pump}_{p+\delta}) t} \hat{a}_{k+\delta}- \mathrm{i} K_{p,p+\delta,k-\delta,k} A^*_p A_{p+\delta}e^{\mathrm{i}\omega^\mathrm{pump}_p t} e^{-\mathrm{i}\omega^\mathrm{pump}_{p+\delta} t} \hat{a}_{k-\delta}-\frac{\kappa_k}{2}\hat{a}_k,\\
\dot {\hat{a}}_{k+i\delta} &=& -\mathrm{i}\omega_{k+i\delta} \hat{a}_{k+i\delta}- \mathrm{i} K_{p,p+\delta,k+i\delta,k+(i-1)\delta} A^*_p A_{p+\delta} e^{\mathrm{i}(\omega^\mathrm{pump}_p-\omega^\mathrm{pump}_{p+\delta}) t}\hat{a}_{k+(i-1)\delta}\\
&&- \mathrm{i} K_{p,p+\delta,k+(i+1)\delta,k+i\delta} A_p A^*_{p+\delta} e^{-\mathrm{i}(\omega^\mathrm{pump}_p-\omega^\mathrm{pump}_{p+\delta}) t}\hat{a}_{k+(i+1)\delta}-\frac{\kappa_{k+i\delta}}{2}{\hat{a}}_{k+i\delta},\qquad i\neq i_\text{max}\\
\dot {\hat{a}}_{k+i_\text{max}\delta} &=& -\mathrm{i}\omega_{k+i_\text{max}\delta} \hat{a}_{k+i_\text{max}\delta}- \mathrm{i} K_{p,p+\delta,k+i_\text{max}\delta,k+(i_\text{max}-1)\delta} A^*_p A_{p+\delta}e^{\mathrm{i}(\omega^\mathrm{pump}_p-\omega^\mathrm{pump}_{p+\delta}) t} \hat{a}_{k+(i_\text{max}-1)\delta}-\frac{\kappa_{k+i_\text{max}\delta}}{2}{\hat{a}}_{k+i_\text{max}\delta}.
\end{eqnarray*}
We introduce the notation 
\begin{equation} \label{lambda-def}
\Delta = \omega^\mathrm{pump}_{p+\delta}-\omega^\mathrm{pump}_p
\end{equation}
 for the energy difference between pumped modes,  and move to the rotating frame
\begin{equation}\label{Eq:b-intro}
\hat{a}_{k\pm i\delta} = \hat{b}_{k\pm i\delta} e^{-i\Omega_{k\pm i\delta} t},\qquad i\in[1,i_\text{max}],
\end{equation}
where the $\Omega$'s have to be chosen such that the equations of motion become time-independent. The readout mode is left unchanged, $\hat{a}_k=\hat{b}_k$.
We observe that the physically motivated choice of the frequency $\Omega_{k\pm i\delta}$ equal to the $i$ times the energy difference between two pumped modes removes the time dependence, 
\begin{equation}
    \Omega_{k\pm i\delta}=\pm i\Delta, \qquad i\in[1,i_\text{max}].
\end{equation}
Using the above expression for $\Omega_{k\pm i\delta}$ and rotating frame~(\ref{Eq:b-intro}), and  substituting $A_p A^*_{p+\delta}=A^*_p A_{p+\delta}=\sqrt{n_p n_{p+\delta}}$ ($n_p$ and $n_{p+\delta}$ are the number of photons in the two pumped modes), we obtain equations without explicit time dependence, 
\begin{align*}
     \dot {\hat{b}}_k & = -\mathrm{i}\omega_k \hat{b}_k - \mathrm{i} K_{p,p+\delta,k,k+\delta} \sqrt{n_p n_{p+\delta}} \,\hat{b}_{k+\delta}- \mathrm{i} K_{p,p+\delta,k-\delta,k}   \sqrt{n_p n_{p+\delta}}\,\hat{b}_{k-\delta}-\frac{\kappa_k}{2}\hat{b}_k,
     \\
  \dot {\hat{b}}_{k+i\delta}& = -\mathrm{i}(\omega_{k+i\delta}-i\Delta) \hat{b}_{k+i\delta}
  - \mathrm{i} K_{p,p+\delta,k+(i-1)\delta,k+i\delta}   \sqrt{n_p n_{p+\delta}}\, \hat{b}_{k+(i-1)\delta}
  \\
  &- \mathrm{i} K_{p,p+\delta,k+(i+1)\delta,k+i\delta}  \sqrt{n_p n_{p+\delta}}\, \hat{b}_{k+(i+1)\delta}-\frac{\kappa_{k+i\delta}}{2}\hat{b}_{k+i\delta},\quad \text{for}\quad i\neq i_\text{max}
  \\
  \dot {\hat{b}}_{k+i_\text{max}\delta}& = -\mathrm{i}(\omega_{k+i_\text{max}\delta}-i_\text{max}\Delta) \hat{b}_{k+i_\text{max}\delta} - \mathrm{i} K_{p,p+\delta,k+i_\text{max}\delta,k+(i_\text{max}-1)\delta} \sqrt{n_p n_{p+\delta}}  \,\hat{b}_{k+(i_\text{max}-1)\delta}-\frac{\kappa_{k+i_\text{max}\delta}}{2}\hat{b}_{k+i_\text{max}\delta}.
\end{align*}
We reported the last three equations, but the strategy is analogous for the first two. The resulting set of all equations can be compactly represented in a matrix form,
\begin{equation}
\frac{d}{dt}\begin{pmatrix}
 {\hat{b}}_{k-i_\text{max}\delta} \\
 {\hat{b}}_{k-(i_\text{max}-1)\delta} \\
      \vdots\\
{\hat{b}}_{k-\delta} \\
 {\hat{b}}_k \\
 {\hat{b}}_{k+\delta}\\
   \vdots\\
{\hat{b}}_{k+(i_\text{max}-1)\delta} \\
 {\hat{b}}_{k+i_\text{max}\delta} \\   
  \end{pmatrix}=-\mathrm{i}\mathbb{A}
\begin{pmatrix}
   {\hat{b}}_{k-i_\text{max}\delta} \\
   {\hat{b}}_{k-(i_\text{max}-1)\delta} \\
   \vdots\\
   \hat{b}_{k-\delta}\\
   {\hat{b}}_k \\
   \hat{b}_{k+\delta}\\
      \vdots\\
     {\hat{b}}_{k+(i_\text{max}-1)\delta} \\
   {\hat{b}}_{k+i_\text{max}\delta} \\   
  \end{pmatrix}.
\end{equation}
The matrix  $\mathbb{A}$ has tridiagonal form with entries 
\begin{align}
\begin{cases}
    &\mathbb{A}[n,n]=\omega_{k+(n-1-i_\text{max})\delta}-(n-1-i_\text{max})\Delta-\mathrm{i}\frac{\kappa_{k+(n-1-i_\text{max})\delta}}{2}\\
    &\mathbb{A}[m,m-1]=\mathbb{A}[m-1,m]=K_{p,p+\delta,k+(m-1-i_\text{max})\delta,k+(m-2-i_\text{max})\delta} \sqrt{n_p n_{p+\delta}} 
\end{cases},
\end{align}
where matrix indices $n$ and $m$ run in the range $n\in[1,2i_\text{max}+1],\,m\in[2,2i_\text{max}+1]$. Moving to the Fourier space, where the time derivative becomes a factor $-i\omega$, and coupling the probed mode $k$ to the incoming radiation via an additional term  $\sqrt{\kappa_{\text{ex},k}}(\hat{a}_\text{IN,R}+\hat{a}_\text{IN,L})$, where  $\hat{a}_\text{IN,R/L}$ is the input field on either side of the cavity, we express $\hat b_k$ via incoming fields as:
\begin{equation}
    \hat{b}_k = \mathrm{i}\sqrt{\kappa_{\text{ex},k}} (\hat{a}_\text{IN,R}+\hat{a}_\text{IN,L}) [(\omega\mathbb{I}-\mathbb{A})^{-1}]_{i_\text{max}+1,i_\text{max}+1}.
\end{equation}
The subscripts select the relevant entry of the inverted matrix $(\omega \mathbb{I}-\mathbb{A})^{-1}$, and $\mathbb{I}$ is the identity matrix of size $(2i_{max}+1)$. Using boundary conditions similar to the previous Section, we finally obtain
\begin{equation}
    S_{21}[\omega]=\mathrm{i} \kappa_{\text{ex},k}\left[(\omega\mathbb{I}-\mathbb{A})^{-1}\right]_{i_\text{max}+1,i_\text{max}+1}.\label{eq:S21theory}
\end{equation}

\subsection*{Application for coupling to both nearest-neighbors}
\label{subsec:nearest_neighbor_s21}
\refeq{eq:S21theory} can be used to obtain a generalization of \refeq{eq:transmission} which takes into account both processes $k\to k+\delta$ and $k\to k-\delta$ (i.e. $i_\mathrm{max}=1$).
This is useful to reproduce experimental data with low pump power, where just two avoided crossings are observed (see \reffigS[b]{2}).
In \refeq{eq:S21theory} we set $i_\mathrm{max}=1$:
\begin{align*}
    S_{21}[\omega]=\mathrm{i} \kappa_{\text{ex},k}\left[(\omega\mathbb{I}-\mathbb{A})^{-1}\right]_{22},
\end{align*}
with
\begin{align*}
	\mathbb{A}=\begin{pmatrix}
	\omega_{k-\delta}+\Delta-\frac{\mathrm{i}\kappa_{k-\delta}}{2} &K_{p,p+\delta,k,k-\delta}\sqrt{n_p n_{p+\delta}}&0\\
	K_{p,p+\delta,k,k-\delta}\sqrt{n_p n_{p+\delta}}&\omega_{k}-\frac{\mathrm{i}\kappa_{k}}{2}&K_{p,p+\delta,k,k+\delta}\sqrt{n_p n_{p+\delta}}\\
	0&K_{p,p+\delta,k,k-\delta}\sqrt{n_p n_{p+\delta}}&\omega_{k+\delta}-\Delta-\frac{\mathrm{i}\kappa_{k+\delta}}{2}
	\end{pmatrix},
\end{align*}
which includes three modes $k-\delta,k,k+\delta$ involved in nearest-neighbor scattering (see \reffigS[b]{2}).
Explicitly,
\begin{align}
	S_{21}[\omega] &= \frac{\kappa_{\text{ex},k}}{\left(\frac{\kappa_k}{2}-\mathrm{i}(\omega-\omega_k)\right)+\frac{K_{p,p+\delta,k,k-\delta}^2n_pn_q}{\frac{\kappa _{k-\delta }}{2}-\mathrm{i}(\omega-\omega _{k-\delta }-\Delta)}+\frac{K_{p,p+\delta,k,k+\delta}^2n_pn_q}{\frac{\kappa _{k+\delta }}{2}-\mathrm{i}(\omega-\omega _{k+\delta }+\Delta)}}=\nonumber \\
	&= \frac{\kex}{\kappa/2 - \mathrm{i}(\omega-\omega_k) + \frac{g^2}{\kappa/2-\mathrm{i}(\omega-\omega_{k-\delta}-\Delta)}+\frac{g^2}{\kappa/2-\mathrm{i}(\omega-\omega_{k+\delta}+\Delta)}},\label{eq:s21_two_anticrossings}
\end{align}
where in the last equation we ignored weak linewidth and matrix element dependence on mode number.

By fitting individual avoided crossings on resonance we extract coupling $g$, $\fany[\prime]{k}$, $\fany[\prime]{k\pm\delta}$ (note that mode frequencies in \refeq{eq:s21_two_anticrossings} are different from undriven ones) and linewidth $\kappa$ (also can be different from undriven case).
If experimental conditions are chosen in a way that $\fany[\prime]{k}$, $\fany[\prime]{k\pm\delta}$ are almost independent on detuning for a certain detuning range, then whole 2D plot can be calculated where the parameters are extracted from two cuts and a good agreement between theory and experiment can be achieved.

\subsection*{Application for cascades}
In case of cascaded scattering \refeq{eq:S21theory} must be used which is trimmed to contain necessary amount of cascades (e.g. for \reffigS[c]{3} we set $i_\mathrm{max}=4$).
Again, we assume that mode frequencies as well as linewidths stay nearly constant through the pump detuning range used in the experiment.
To decrease the number of fitting parameters we introduced a common scaling factor (due to Kerr effect) for mode frequencies which is a single number for all modes participating in cascaded scattering.
Here, instead of fitting the coupling $g$, we kept the matrix element dependence on mode number, see \refeq{eqn:boxmatrixel}, and fitted the factor between $g$ and square root of product of mode numbers participating in particular process.
This factor corresponds to either fitting of $\Eg$ or occupation of pumps (due to systematic error in cryostat insertion loss).
Thus in total we have 5 fitting parameters: linewidth ($\kappa_\text{fit}$), frequency correction due to Kerr ($\alpha_\text{fit}$), occupation of the pumps $n^\text{pump}_\text{fit}=\sqrt{n_p n_{q}}$, and magnitude ($S_\text{fit}$) and phase offset ($\varphi_\text{fit}$) coming from imperfect calibration of the cryostat.
Explicitly, our fitting function is:
\begin{equation}
    S_{21}[\omega]=\mathrm{i}S_\text{fit} \kappa_{\text{ex},k}\left[(\omega\mathbb{I}-\mathbb{A}_\text{fit})^{-1}\right]_{i_\text{max}+1,i_\text{max}+1}e^{\mathrm{i}\varphi_\text{fit}},
	\label{eq:s21_cascades_generic}
\end{equation}
with
\begin{align}
\begin{cases}
    &\mathbb{A}_\text{fit}[n,n]=\omega_{k+(n-1-i_\text{max})\delta}(1+\alpha_\text{fit})-(n-1-i_\text{max})\Delta-\mathrm{i}\frac{\kappa_\text{fit}}{2}\\
    &\mathbb{A}_\text{fit}[m,m-1]=\mathbb{A}[m-1,m]=K_{p,p+\delta,k+(m-1-i_\text{max})\delta,k+(m-2-i_\text{max})\delta}n^\text{pump}_\text{fit}.
\end{cases}
\end{align}

\section*{Many-mode drives problem}
\label{sec:Kinetic_equation}
In this Section we provide the main points of our treatment of the multimode drive problem: the kinetic equation and the self-consistent approach to the linewidth. Although a rigorous description of such problem requires setting a master/Lindblad formalism, for our phenomenological description we assume that the state of the system is described solely by the occupation numbers, $n_k = \langle \hat{a}^\dagger_k\hat{a}_k \rangle$. The equilibration of the driven JJ chain results from the interplay of two processes: the thermal relaxation process, that aims to relax the occupation numbers $n_k$ to thermal equilibrium at the base temperature, and the external driving that aims to increase occupation of pumped modes. In the process of equilibration, the role of the intrinsic scattering between modes from nonlinearity depends on the mode occupation and strength of the driving.

\subsection*{From a single driven mode to many driven and interacting modes}
We start with the general form of the kinetic equation of a single plasmonic mode coupled to a background bath with coupling $\kappa_k$ and whose population is replenished by an external pump-tone:
\begin{equation}
 \dot n_k = - \kappa_k (n_k+1)n_k^\text{th} +\kappa_k n_k (n_k^\text{th}+1) + \kappa_{\text{ex},k} n^\text{flux}_k = - \kappa_k (n_k-n_k^\text{th}) + \kappa_{\text{ex},k} n^\text{flux}_k.\label{eqn:firstkin}
\end{equation}
Here $n_k^\text{th} = (e^{\hbar \omega_k/(k_BT)}-1)^{-1}$ is the thermal occupation of mode $k$ according to Bose-Einstein distribution at temperature $T$,  $\kappa_k$ is the total coupling rate to the thermal bath (including coupling to both terminals and to the internal environment), and $\kappa_{\text{ex},k}$ is the coupling rate to each of the external terminals. When $n^\text{flux}_k$ --- the number of injected photons per unit frequency and unit time --- is zero, this equation yields the expected exponential in time relaxation of $n_k$ to the thermal occupation. Otherwise, the population of the given mode is increased by an amount set by the ratio of $\kappa_{\text{ex},k}$ and $\kappa_k$. The nonequilibrium steady state (NESS) is obtained as a stationary point
\begin{equation}
    \dot{n}_k=0\Rightarrow n_k = n_k^\text{th}+ \frac{\kappa_{\text{ex},k}}{\kappa_k} n^\text{flux}_k,
\end{equation}
and is characterized by the excess occupation proportional to the external photon flux. 

We generalize \refeq{eqn:firstkin} to the multi-mode case, where all modes interact with each other and multiple modes are pumped. In the experiment, the photon flux is non-zero for the lowest 19 modes, and is related to the noise power spectral density $P$ at the input of the cavity as $n^\text{flux}_k=P/(h\omega_k)$. Besides that, the essential missing contribution to the kinetic equation~(\ref{eqn:firstkin}) are inter-mode scattering processes caused by the nonlinearity, that conserve both energy and momentum. These processes are accounted by the collision integral, that  depends on the occupation factors of all modes and on the transition probability arising from $H_\text{int}$, calculated with a Fermi golden rule approach. For each scattering process involving a set of modes $k,p,q_1,q_2$, the collision integral consists of an in- and an out-scattering component~\cite{lin_prl_2013}: 
\begin{equation}\label{Eq:collision}
I_\text{in}[k] 
=
\sum_p \sum_{q_1>q_2} W_{q_1q_2\to p,k} (1+n_p) (1+n_k) n_{q_1} n_{q_2}
,
\quad
I_\text{out}[k] 
=
- \sum_p \sum_{q_1>q_2} W_{ p,k \to q_1q_2} n_p n_k (1+n_{q_1}) (1+n_{q_2}),
\end{equation}
where $W$ depends on the matrix element as (${\hbar}=1$)
\begin{equation}
 W_{q_1q_2\to p,k}
 =
  W_{ p,k\to q_1q_2}
 =
2\pi |K_{q_1q_2pk}|^2 \delta(\omega_{q_1}+\omega_{q_2}-\omega_p-\omega_k).
 \label{eqn:Wrate}
\end{equation}
In the above sums, we exclude by hand the unphysical cases of self-decay, i.e. $q_1=k,\,q_2=k$.

Finally, the full kinetic equation including collision integral is written as:
\begin{equation}
    \dot n_k = - \kappa_k (n_k-n_k^\text{th}) +\kappa_{\text{ex},k} n^\text{flux}_k+ I_\text{in}[k]+ I_\text{out}[k],
    \label{Eq:kin}
\end{equation}
where we use a set of modes up to a certain cutoff, $k=1,\dots,k_\text{max}$. 

\subsection*{Self-consistent linewidth and final excess linewidth}
The transition probability~(\ref{eqn:Wrate}) contains a delta-function in energy. In practice, every mode has a Lorentzian shape with a finite linewidth. Hence, we replace the $\delta$-function by a broadened form, $\delta(\omega) \to    \delta_\gamma(\omega)$, 
\begin{eqnarray}\label{Eq:broaden}
    \delta_\gamma(\omega) = \frac{1}{\pi}\frac{\gamma}{\gamma^2+\omega^2},
\end{eqnarray}
where $2\gamma$ physically corresponds to the total linewidth. Since four convolved Lorentzians produce a Lorentzian with linewidth given by the sum of the original linewidths, we replace the $\delta$ function in \refeq{eqn:Wrate} by the expression \refeq{Eq:broaden} above with broadening $\gamma$ given by:
\begin{equation}\label{gamma4}
\gamma_{k,p,q_1,q_2} = \sum_{j \in [k,p,q_1,q_2]} (2\kappa_{\text{ex},j}+\kappa_{\text{i},j}+\delta\kappa_j)/2,
\end{equation}
which depends symmetrically on the four mode numbers that participate in the scattering process (this is required to fulfill the detailed balance). The first two terms in the sum are originating from external broadening and internal losses. The third contribution to the sum in \refeq{gamma4}, $\delta \kappa_j$, is the (self-consistent) excess linewidth of mode $j$ due to the intrinsic scattering caused by the nonlinearity. 

Within the framework of kinetic equation, the excess intrinsic linewidth is obtained from the diagonal part of the linearized collision integral,
 \begin{equation}\label{Eq:decay-rate}
 \delta\kappa_k 
 =
\frac{\pi^5 \Eg^2}{16N^6} \sum_{p,q_1,q_2} kpq_1q_2 \,\delta(\omega_k+\omega_p-\omega_{q_1}-\omega_{q_2}) [n_p(1+n_{q_1}+n_{q_2})-n_{q_1}n_{q_2}]\sum_{s_1,s_2,s_3=\pm}\delta_{k+s_1p+s_2{q_1}+s_3{q_2},0},
\end{equation}
where again the first sum excludes self-decays, and we used explicit expression for the matrix element from \refeq{eqn:boxmatrixel}.

\subsection*{Obtaining NESS and linewidth numerically}
In order to model the increase of the linewidth in observed in the experimental data, we calculate the NESS of the kinetic equation with fixed $n^\text{flux}_k$. In the simulations we used $k_\text{max} = 173$ and smaller values $k_\text{max} = 100$ to make sure that occupation of  modes with numbers much below $k_\text{max}$ are insensitive to precise value of the $k_\text{max}$.  We get the NESS by time evolving the distribution function according to \refeq{Eq:kin} until the norm of the change of $n_k$ between two consecutive time steps becomes smaller than tolerance, $\sum_k |\delta n_k|^2 \leq 10^{-14}$. During the time evolution with time step $\Delta t = 0.01/\kappa_0$ ($\kappa_0=1.7$~MHz is an average value for $\kappa_k$), we update the value of $\delta \kappa_j$ using \refeq{Eq:decay-rate} at every $10^\text{th}$ time step. After the convergence is reached, we use the converged values of the distribution function in the NESS, $n_k$, to calculate $ \delta\kappa_j $ that is plotted in \reffigS[c]{5}.

Although the system is fully described by the three energy parameters, achieving agreement between simulations and experiments required introducing two \emph{additional} phenomenological fitting parameters, resulting in the modified kinetic equation:
\begin{equation}
    \dot n_k = - (\kappa_k+\delta\kappa_{\mathrm{i},k}) (n_k-n_k^\text{th}) 
    + \alpha\kappa_{\text{ex},k}  n^\text{flux}_k
    + I_\text{in}[k]+ I_\text{out}[k].
\end{equation}
In this equation the following parameters are introduced:
\begin{enumerate}
\item The first parameter is a frequency-independent correction to the insertion loss of the fridge, $\alpha$, which affects the magnitude of the term proportional to $n^\text{flux}_k$.
We set $\alpha=5$ based on comparisons between the 20~dB data and simulations.
As point of comparison, the measured nonlinearity in $K/2\pi = 16$~kHz/ph from~\reffigS[e]{2}, indicates an insertion-loss correction of 3.
The insertion-loss correction determined from nonlinearities thus indicates that our independent throughline insertion loss measurements underestimate the total insertion loss by $5-7~\mathrm{dB}$.

\item For stronger drives, we introduce an additional internal loss, $\delta\kappa_{\mathrm{i},k}$, for the driven modes with numbers $k=1,\ldots,19$,
Setting $\delta\kappa_{\mathrm{i},k}=6$~MHz for the 5~dB data results in a good match between theory and experiment.
The introduction of $\delta\kappa_{\mathrm{i},k}$ is justified by the evident presence of additional losses in the lowest measured modes at strong drives (e.g., see the 0~dB data in \reffigS[c]{5}).
This excess loss might be attributed to phase slips, which are beyond the scope of our theoretical model.
\end{enumerate}

\subsection*{Estimate of the number of decay channels}
Lastly, we comment on the number of decay channels arising in the multimode drive case for a mode that is not being driven. As an example, we consider mode $k=80$. In order to visualize the number of decay channels we first obtain numerically NESS. After this, we plot the contribution of individual terms in the \refeq{Eq:decay-rate} to the linewidth of the $k$-th mode. The decay processes of mode $k$ are labeled by three numbers, $k,p\to q_1,q_2$, however the number $p$ is deduced from momentum conservation, leaving only two independent mode numbers, $q_1$ and $q_2$. In order to avoid dealing with multiple signs of momenta ($s_{1,2,3}=\pm1$ in \refeq{Eq:decay-rate}), we plot the contribution to $\delta \kappa_k$ as a function of $s_2 q_1$ and $s_3 q_2$, so that they take both positive and negative values now. The total excess linewidth results from the sum of all matrix elements, but here we want to get an insight into the mostly contributing decay channels.

First we show the matrix plot in Fig.~\ref{fig:decay-estimate}A where we mark by black color the values of $(s_2 q_1,s_3 q_2)$ that exhaust the 95\% of the total linewidth. From here we see that relevant processes are of two types. The first is scattering to nearby modes (i.e. when either $q_1$ or $q_2$ is in the vicinity of $k=80$), which corresponds to the up-right and bottom-left features in the matrix plot. This scattering is shown schematically in \reffigS[a]{5} of the main text. The second scattering process is reminiscent of the scattering process discussed in \cite{bard_prb_2018} for an infinite Josephson junction chain at thermal equilibrium, where $k$ relaxes by scattering off a plasmon with much smaller mode number $p$ and negative momentum. In our matrix plot, this corresponds to the black entries following a diagonal trend in \reffig[a]{fig:decay-estimate}.

In addition, in \reffig[b]{fig:decay-estimate} we show the relative cumulative rate, obtained by progressively summing contributions of all decay channels, from the largest to the smallest one, normalized by total decay rate. This can be used to visualize the \emph{number} of decay channels for mode $k$. Once the number is divided by two, to account for the indistinguishability of $q_1$ and $q_2$, we obtain an estimate of a bit more than 100 channels to achieve a cumulative proportion of 95\%.

\section*{Estimation of scattered photons}
\label{sec:estimation_noise}
The noise data in \reffigS[b]{4} (see also \reffig{fig:sup_noise_cascades} for a different pumping configuration) can be further analyzed quantitatively.
For that it is necessary to calibrate the measurement chain and determine the amplifier-referred gain, $G$, and added noise (see \reffig{fig:sup_calibration}).
The background in \reffigS[b]{4} and \reffig{fig:sup_noise_cascades} is set by the added noise within a 270~Hz band, which was measured to be approximately $-103$~dBm for the frequency range used in those measurements.
Regions where $P_\mathrm{N}$ significantly exceeds the background can be interpreted as cavity emission at a rate proportional to $\kex$.
Using an independent calibration of the input-referred gain, $G$, the number of photons inside the mode can be estimated as $\frac{P_\mathrm{N}}{G\kex}\frac{1}{h\fany{\mathrm{N}}}$.
For $P_\mathrm{N} = -75$~dBm, this estimate yields approximately half a photon for a typical $\kex = 2\pi \times 0.4$~MHz, which agrees well with the estimation for resonant scattering $0.5 - 1$, given by $n_k\cdot g/\kappa$.


\begin{figure}
	\centering\includegraphics[width=5.9in]{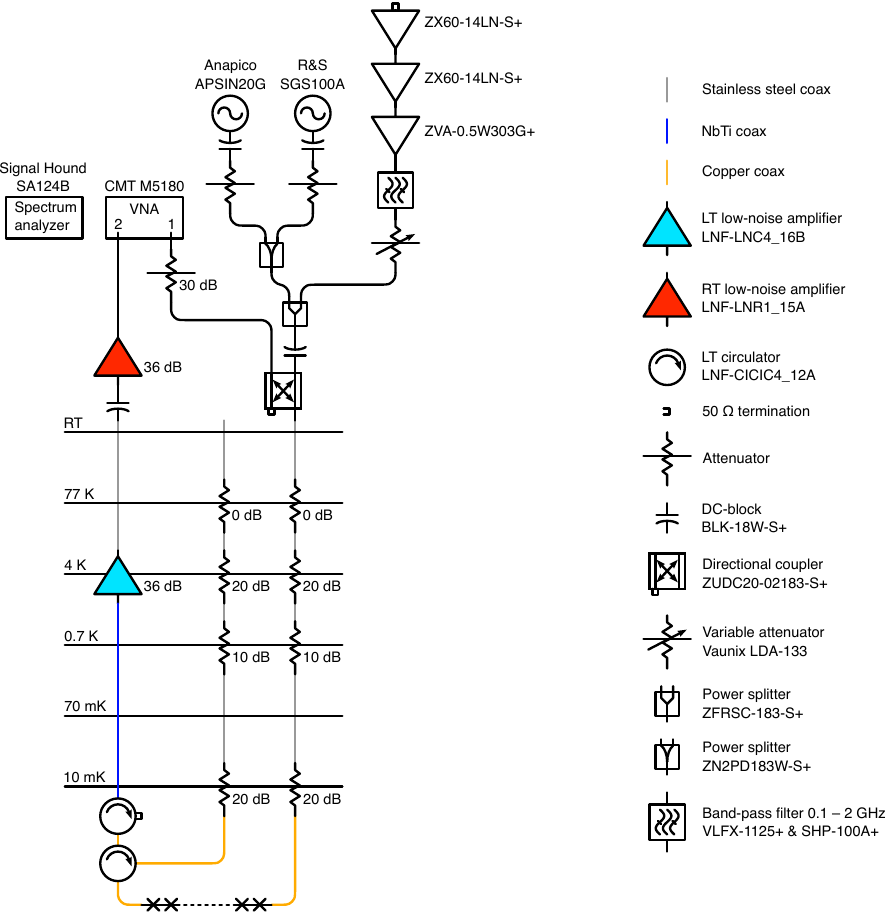}
	\caption{%
	\textbf{Experimental setup used in this work.}
	Power splitters and directional coupler combines microwave radiation from all sources and feeds it into the fridge.
	Passed though the series of cold attenuators and the device the signal is amplified by a cascade of low-temperature and room-temperature amplifiers and is read-out by vector network analyzer (VNA).
	Alternatively, the signal out-coming from the fridge can be detected by the spectrum analyzer.
	A cascade of RT amplifiers plays role of a wide band noise source (the band is defined by a series of low-pass and high-pass filters) where the variable attenuator allows us to control the noise temperature at the input of cryostat roughly in a range of $300 - 3\cdot10^8$~K.%
	}
	\label{fig:fridge}
\end{figure}
\clearpage

\begin{figure}
	\centering\includegraphics[width=5.0in]{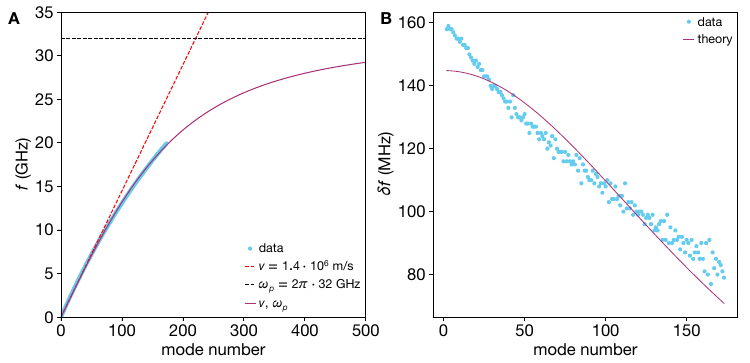}
	\caption{%
	\textbf{Dispersion curve of the JJ-chain.}
	\textbf{(A)}~Blue data points represent the frequencies of the first 173 modes measured using the two-tone spectroscopy technique (see details in the text).
	The blue solid line corresponds to a fit to \refeq{eq:dispersion}, while the two dashed lines indicate two relevant quantities: the speed of light, $v$, and the plasma frequency of a single junction, $\omega_\mathrm{p}$. \textbf{(B)}~Mode spacing.
	Blue data points represent experimental data (the same as in~\textbf{(A)}).
	The theoretical curve is obtained by differentiating the fit from panel \textbf{(A)}.
	}
	\label{fig:dispersion}
\end{figure}
\clearpage

\begin{figure}
	\centering\includegraphics[width=6.00in]{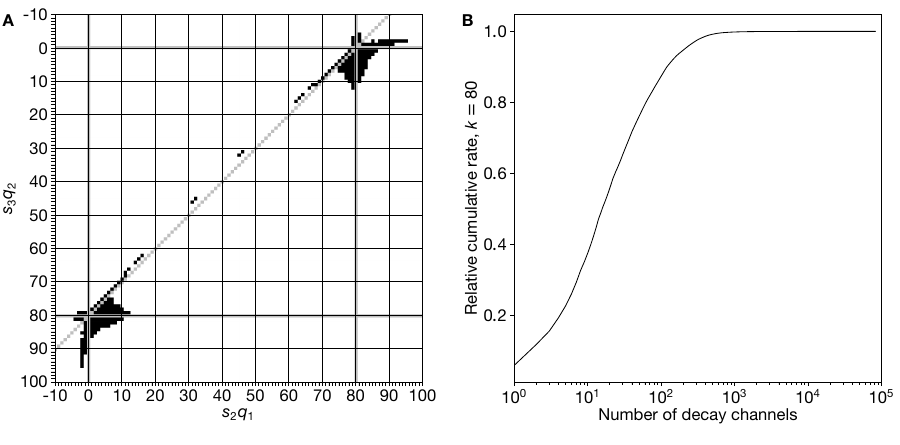}
	\caption{%
	\textbf{Analysis of decay channels for a not driven mode.}
	\textbf{A.}~Matrix plot of contributions to the decay rate of mode $k=80$. Each entry corresponds to a different decay channel, identified by $s_2 q_1$ and $s_3 q_2$ in \refeq{Eq:decay-rate}. Gray lines indicate the unphysical cases of self-decay process (either $q_1$ or $q_2$ being equal to $k$), or mode numbers $q_1,\,q_2,\,p=0$, which is just a numerical artifice. \textbf{B.}~Relative cumulative rate for the same mode $k=80$, which allows to easily estimate the number of decay channels needed to accumulate a given percentage of the total rate.
	}
	\label{fig:decay-estimate}
\end{figure}
\clearpage

\begin{figure}[!ht]
	\centering\includegraphics[width=7.24in]{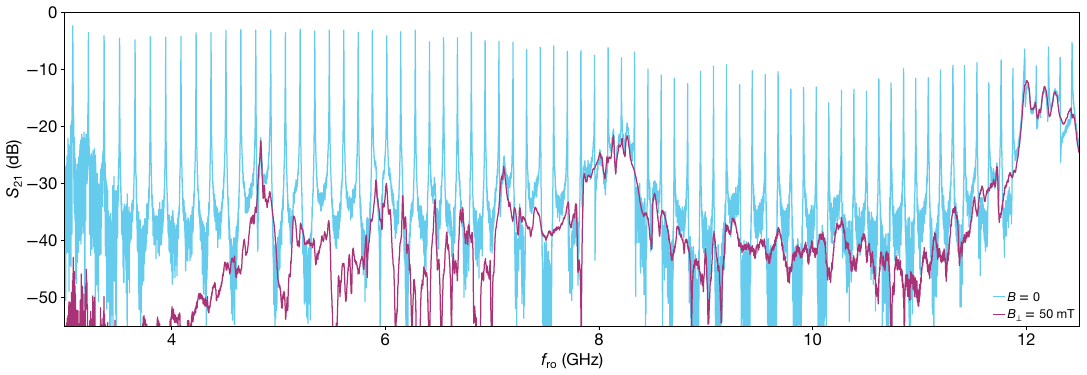}
	\caption{%
	\textbf{Single tone data without background subtraction.}
	Single-tone transmission data, $S_{21}$, is presented at zero magnetic field and at $B_\perp = 50$~mT, where the JJ-chain transitions to normal state.
	The normal state data serves as a background representing cross-talk between the two ports of the printed circuit board, which can be subtracted (in linear voltage ratio units).
	In \reffigS[b]{1} of the main text, the $S_{21}$ data was subtracted for visualization purposes, whereas in the remaining figures, the unsubtracted data is reported.
	Data within frequency ranges exhibiting high cross-talk (e.g., at $\fro \sim 8$~GHz) were compared with the $B_\perp = 50$~mT trace and excluded from the analysis if the subtraction procedure caused significant changes in the fitted resonance parameters (e.g., low attenuation data in \reffigS[c]{5} for $k \approx 60$).
	}
	\label{fig:sup_subtraction}
\end{figure}
\clearpage

\begin{figure}
	\centering\includegraphics[width=5.00in]{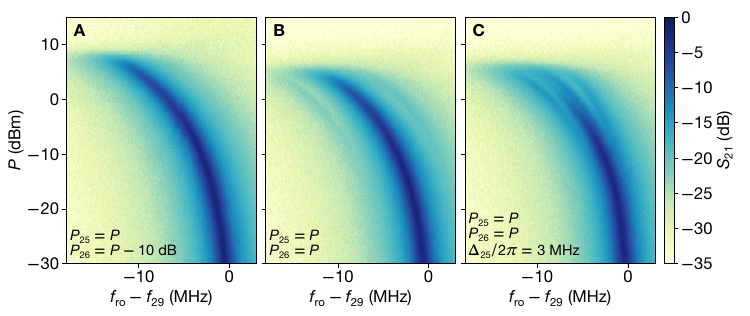}
	\caption{%
	\textbf{Pump imbalance and pump detuning effects.}  
	Three-tone spectroscopy data for the similar configuration of pumps ($p=25$, $q=26$) and read-out ($k=29$) as in \reffigS[e]{1}.  
	\textbf{A.}~Pump powers are imbalanced, with $P_{25}$ higher than $P_{26}$ by 10~dB.  
	\textbf{B.}~Pump powers are equal, $P_{25} = P_{26} = P$.  
	\textbf{C.}~Pump powers are equal, but the pump frequency of mode 25 is blue-detuned by 3~MHz compared to panels (\textbf{A, B}).  
	This is the same data as presented in \reffigS[e]{1}.  
	}
	\label{fig:pump_imbalance}
\end{figure}
\clearpage

\begin{figure}
	\centering\includegraphics[width=7.24in]{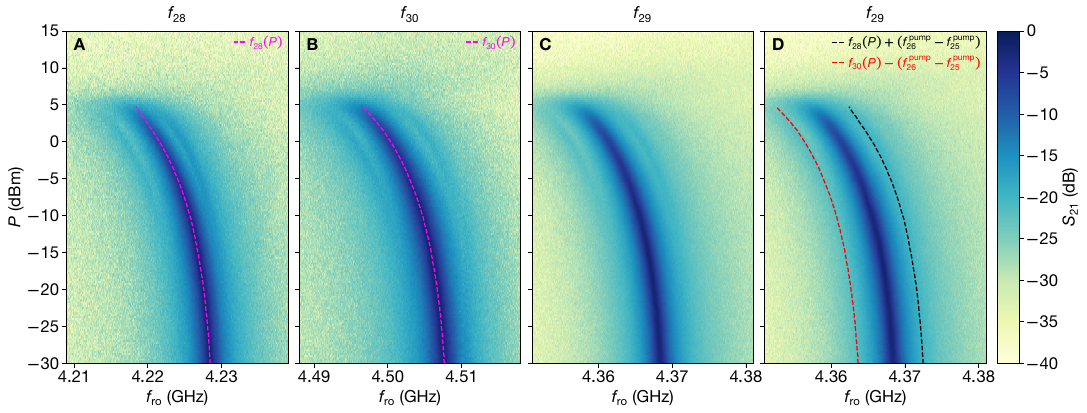}
	\caption{%
	\textbf{Sideband feature energetics}  
	Three-tone spectroscopy data for the same configuration of pumps ($p=25$, $q=26$) as in \reffig[b]{fig:pump_imbalance}.
	In all panels, the pumping configuration was kept identical, and only the read-out tone ($k$) was varied.
	\textbf{A-B.}~$S_{21}$ measurements around modes 28 and 30.
	The resonances $\fany{28}(P)$ and $\fany{30}(P)$ were determined as functions of the pump power $P$.
	\textbf{C-D.}~The same data as in \reffig[b]{fig:pump_imbalance}.
	In panel \textbf{D}, two guides are overlaid, showing predictions for a four-wave mixing-like interaction based on a quasi-classical energy conservation argument.  
	}
	\label{fig:sideband_energetics}
\end{figure}
\clearpage

\begin{figure}
	\centering\includegraphics[width=7.24in]{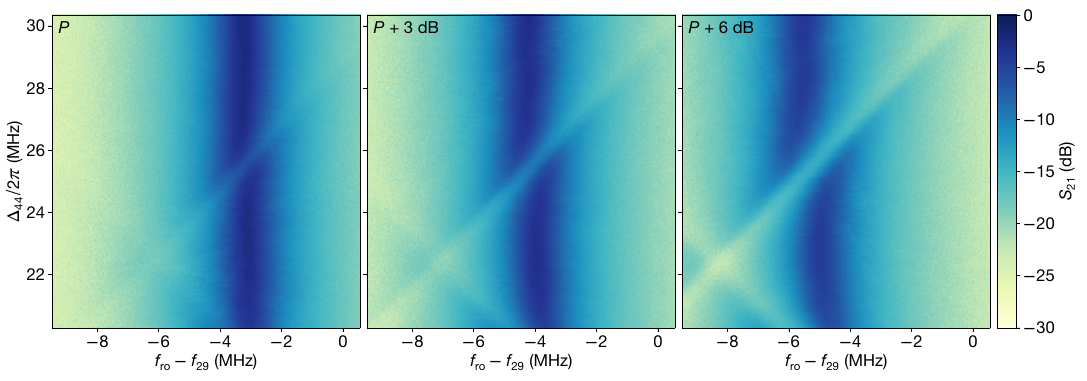}
	\caption{%
	\textbf{Avoided crossing power dependence}  
	$S_{21}$ data near the avoided crossing for a similar pumping configuration as in \reffigS[b]{2}.  
	A zoomed-in view of the top avoided crossing is shown in each panel.  
	The pump power for both pump tones was varied simultaneously, increasing by 3~dB in each panel. 
	Data in \reffigS[d]{2} correspond to cuts at the avoided crossing from this dataset.  
	}
	\label{fig:avoided_power_dependence}
\end{figure}
\clearpage

\begin{figure}
	\centering\includegraphics[width=7.24in]{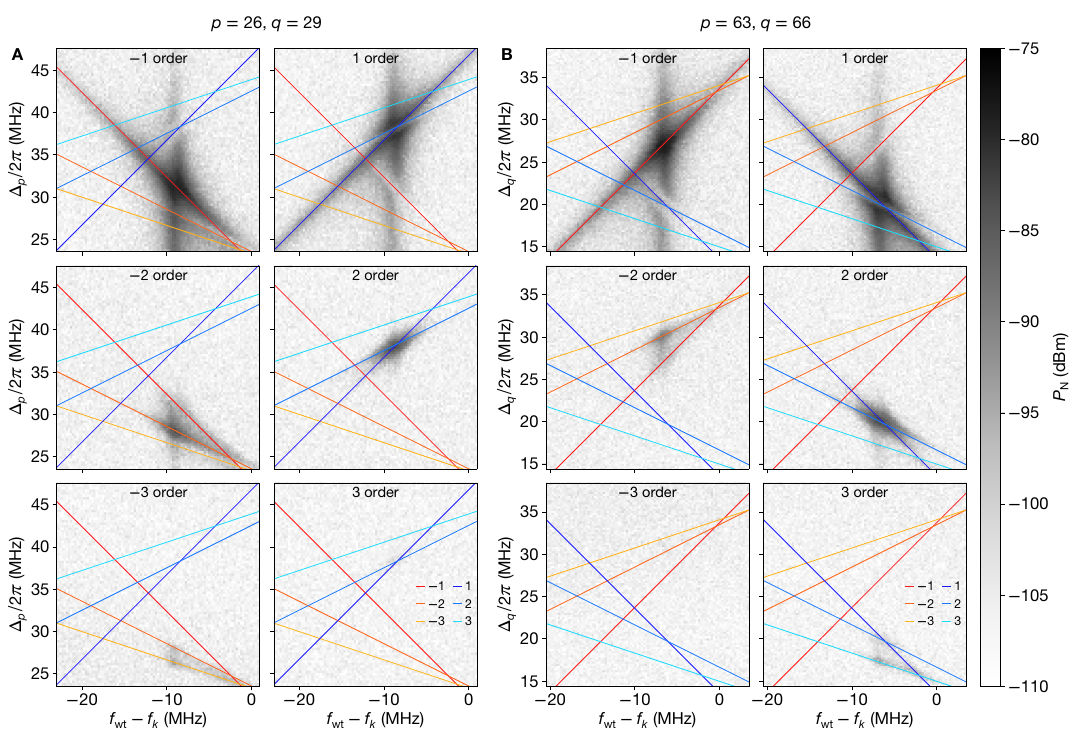}
	\caption{
	\textbf{Observation of scattered photons via noise measurements}
	Noise measurement data for a different pumping configuration with $p, q < k$.
	In~\textbf{B}, the data from \reffigS[b]{4} is replotted.
	A weak tone is applied at $k=46$ in both panels, and the pump powers for \textbf{A} were chosen to result in a similar matrix element as the pumping configuration in \textbf{B}.
	}
	\label{fig:sup_noise_cascades}
\end{figure}
\clearpage

\begin{figure}
	\centering\includegraphics[width=5.00in]{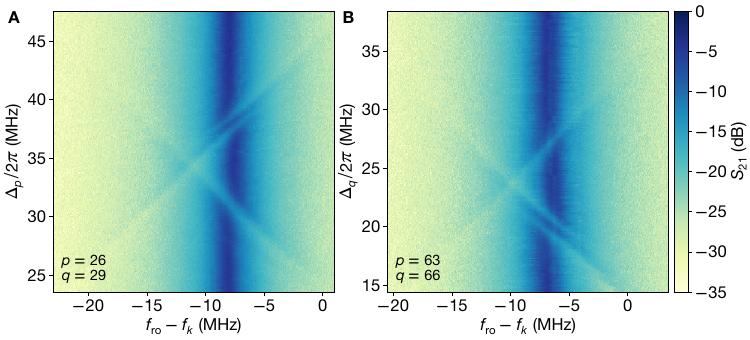}
	\caption{
	\textbf{Moderate pump power three-tone spectroscopy data}
	\textbf{A.}~Transmission data measured under the same pumping configuration as the noise data in \reffig[a]{fig:sup_noise_cascades}.
	\textbf{B.}~Transmission data measured under the same pumping configuration as the noise data in \reffigS[b]{4} and \reffig[b]{fig:sup_noise_cascades}.
	}
	\label{fig:sup_s21_cascades}
\end{figure}
\clearpage

\begin{figure}
	\centering\includegraphics[width=6.00in]{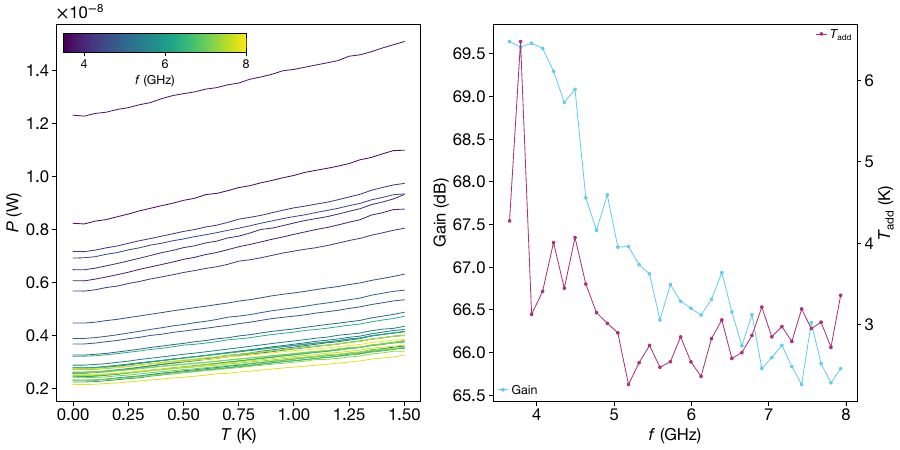}
	\caption{%
	\textbf{Calibration of measurement chain of the setup}  
	The power measured by a spectrum analyzer, $P$, emitted from the cryostat, is shown for different center frequencies (denoted by color) as a function of the mixing chamber plate temperature, $T$.  
	All data were recorded in 15~MHz bands centered at the same frequencies as the data in \reffigS[d-e]{5}, to calibrate the input-referred gain, $G$, and the added noise temperature, $T_\mathrm{add}$.  
	The gain and added noise temperature were extracted from a linear fit of $P$ versus $T$ using the relation $P = G(T + T_\mathrm{add})$ for $0.25 \leq T \leq 1.4$~K.  
	}
	\label{fig:sup_calibration}
\end{figure}
\clearpage

\begin{figure}
	\centering\includegraphics[width=7.24in]{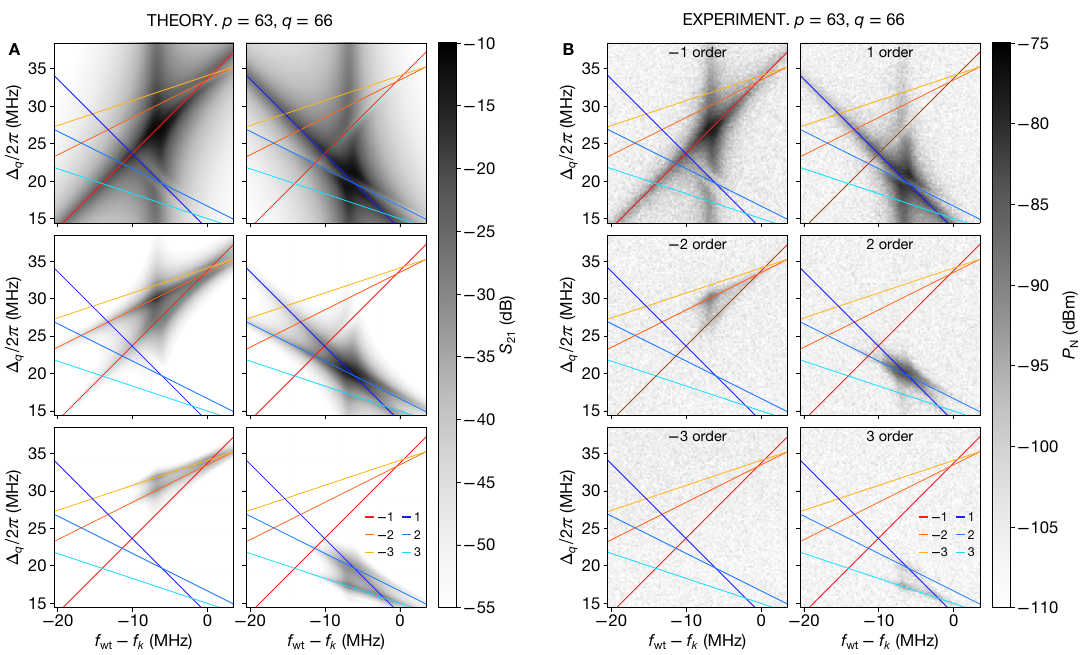}
	\caption{
	\textbf{Comparison of theory and experiment for cascades}
	\textbf{A.}~Theoretical calculation of off-diagonal scattering parameter $S_{21}[k \rightarrow k+i\delta]$ between mode $k$ and $k + i\delta$ using \refeq{eq:s21_cascades_generic}.
	The panel is plotted using matrix element fitted from \reffig[b]{fig:sup_s21_cascades} which is measured at the same pumping configuration as panel \textbf{B}.
	In~\textbf{B}, the data from \reffigS[b]{4} is replotted.
	}
	\label{fig:noise_cascades_theory}
\end{figure}
\clearpage

\begin{table}
	\caption{\textbf{Estimation of JJ-chain parameters.}
	The data in column `Value' are obtained from dispersion fit and geometry of the transmission line. In `Consistency check' the data are calculated from dispersion fit and \textit{assuming} $50-100~\mathrm{fF/\mu m^2}$ of specific capacitance~\cite{kerr_ieee-microwave_1992,deppe_determination_2004}.
	For each column the theoretically calculated parameter is marked with an asterisk.} 
	\centering\begin{tabular}{|c|c|c|c|} 
	\hline
	Parameter & Value & Consistency check \\
	\hline
	$\Eg/h$ & 3.9~THz$^*$ & 2.9\,--\,5.8~THz \\ 
	$\Ej/h$ & 47~GHz & 32\,--\,64~GHz  \\
	$\Ec/h$ & 11~GHz & 8\,--\,16~GHz$^*$ \\
	$Z$ & 13~k$\Omega$ & 10\,--\,20~k$\Omega$ \\
	\hline
	\end{tabular}
	\label{table:parameters}
\end{table}
\nocite{kuzmin_natphys_2019, masluk_prl_2012, mukhopadhyay_natphys_2023,gevorgian_ieee-microwave_1995, simons_book_2001, kerr_ieee-microwave_1992,deppe_determination_2004, bard_prb_2018, lin_prl_2013}
\end{document}